\documentclass[%
 reprint,
 amsmath,amssymb,
 aps,
]{revtex4-2}

\usepackage[utf8]{inputenc}

\usepackage{amsmath,amsfonts,amssymb,bm,cancel,multirow,dsfont}
\usepackage{graphicx,float} 
\usepackage{enumitem}
\usepackage{color}
\usepackage{latexsym}
\usepackage{wasysym}
\usepackage{soul}
\usepackage{dcolumn}
\usepackage{xcolor}
\usepackage{placeins}
\usepackage{ bbold }
\usepackage[final]{changes}

\usetikzlibrary{arrows.meta}
\newcommand{\dup}{\ensuremath{\mathrm d}}

\newcommand{\ld}[2]{\frac{\dup #1}{\dup #2}}

\newcommand{\erf}{\text{erf}}

\newcommand{\tens}[1]{ \underline{\underline{#1}}}
\newcommand{\tr}{\text{tr}}
\newcommand{\vc}[1]{\underline{#1}}
\newcommand{\xm}{\langle x_{\min} \rangle}
\newcommand{\xdotc}{\dot{x}_{\text{c}}}
\newcommand{\twait}{\langle t_{\text{load}} \rangle}
\newcommand{\xdotover}{ \dot{x}_{\text{overlap}}}

\definechangesauthor[name=Daniel,color=blue]{dk}

\newcommand{\dkadd}[1]{\added[]{#1}}
\newcommand{\dkdel}[1]{\deleted[]{#1}}
\newcommand{\dkrep}[2]{\replaced[]{#1}{#2}}

\begin{document}
\title{Dynamic phase diagram of plastically deformed amorphous solids at finite temperature}

\author{Daniel Korchinski}
\email{djkorchi@phas.ubc.ca}
\affiliation{
 Department of Physics and Astronomy and Stewart Blusson Quantum Matter Institute,  University of British Columbia, Vancouver BC V6T 1Z1, Canada}%
\author{J\"org Rottler}%
\affiliation{
 Department of Physics and Astronomy and Stewart Blusson Quantum Matter Institute,  University of British Columbia, Vancouver BC V6T 1Z1, Canada}%

\begin{abstract}
    The yielding transition that occurs in amorphous solids under athermal quasistatic deformation has been the subject of many theoretical and computational studies. Here, we extend this analysis to include thermal effects at finite shear rate, focusing on how temperature alters avalanches. We derive a nonequilibrium phase diagram capturing how temperature and strain rate effects compete, when avalanches overlap, and whether finite-size effects dominate over temperature effects. The predictions are tested through simulations of an elastoplastic model in two dimensions and in a mean-field approximation. We find a new scaling for temperature-dependent softening in the low-strain rate regime when avalanches do not overlap, and a temperature-dependent Herschel-Bulkley exponent in the high strain rate regime when avalanches do overlap. 
\end{abstract}

\maketitle

\section{Introduction}
Amorphous solids are materials that, like fluids, lack long-range order on the constituent particle scale, yet are solid at rest. When deformed slowly, these solids respond, first \dkadd{seemingly} elastically and then once a critical stress is achieved, plastically~\cite{bonn_yield_2017,schuh_mechanical_2007}. This response can be either brittle, with system spanning shear bands, or ductile, with homogeneous plasticity~\cite{schuh_mechanical_2007,manning_rate-dependent_2009,fielding_shear_2014,barlow_ductile_2020}. Ductile plastic flow is ``jerky'', with periods of elastic stress-loading punctuated by short bursts of stress-releasing plastic rearrangement \dkadd{dubbed ``avalanches''}. These \dkrep{avalanches}{rearrangements} can be decomposed into individual shear-transformations (STs), regions of plastic deformation typically involving a few tens of particles \cite{argon_plastic_1979,argon_plastic_1979-bubble,maloney_amorphous_2006}. Each ST causes a rearrangement that locally relieves stress, while inducing a long-range a quadrupolar stress-field \cite{maloney_amorphous_2006,nicolas_elastic_2015,albaret2016mapping,nicolas_orientation_2018} that can trigger further STs in an avalanche. In the flowing state, in absence of flow inhomogeneities, the hallmarks of a dynamical phase transition emerge: avalanches are scale-free, with non-trivial critical exponents \cite{lin_scaling_2014,sandfeld_avalanches_2015,jagla_avalanche-size_2015,liu_driving_2016,lin_mean-field_2016,budrikis_universal_2017,fernandez_aguirre_critical_2018,ferrero_criticality_2019,ferrero_properties_2021}.

Much theoretical attention has been paid to the ductile yielding  transition in the athermal and quasistatic (AQS) limit ~\cite{lin_density_2014,lin_scaling_2014,lin_mean-field_2016,ferrero_criticality_2019,ferrero_properties_2021,le_goff_criticality_2019}. This limit is appropriate when the plastic ST timescale $\tau$, over which rearrangements occur, is much smaller than the periods of elastic loading (set by the driving rate) and the timescale of thermally triggered STs. When driving rate competes with the timescale of plastic rearrangements, rheological effects begin to alter the phase-transition picture~\cite{karimi_inertia_2017,salerno_avalanches_2012,nicolas_effects_2016,salerno_effect_2013}. For systems where the constituent particles are large and Brownian motion small (e.g. foams, emulsions, dense suspensions, etc.) thermal effects can be safely neglected~\cite{nicolas_deformation_2018}. However, for systems with smaller particles or higher temperatures (e.g. metallic glasses close to the glass-transition temperature $T_g$), thermal effects begin to play a role~\cite{schuh_mechanical_2007}.

Thermal history \dkadd{and preparation} can affect whether yielding is  brittle or ductile~\cite{schuh_mechanical_2007,homer_mesoscale_2009,shi_strain_2005,shi_atomic-scale_2006,barlow_ductile_2020}.
Specifically in ductile steady-state flow, the competition of thermal and driving-rate timescales \dkrep{changes the rheology}{can bring new rheological effects into play,} \dkadd{as follows: }
If driving rates $\dot{\gamma}$ are increased, so that the loading time between avalanches becomes so short as to be comparable to the duration of the average  avalanche $\langle t\rangle_{\text{av}}$, avalanches begin to overlap \dkadd{temporally}, as shown for instance in the molecular dynamics study of Karmarkar et al.~\cite{karmakar_statistical_2010}. 
These overlapping avalanches destroy the anomalous stress-fluctuations of the AQS yielding transition, and the flow stress $\Sigma$ rises above the athermal critical stress $\Sigma_c$ according to the Herschel-Bulkley law~\cite{becu_yielding_2006,lin_scaling_2014,caggioni_variations_2020},
\begin{equation}
\Sigma(T=0,\dot{\gamma}) = \Sigma_c + A \dot{\gamma}^{n} \,, \label{eq:herschel_bulkely_defn} 
\end{equation}
\dkadd{ where $n$ is the Herschel-Bulkley exponent and $T=0$ indicates the athermal limit. } \dkadd{ If eq.~\ref{eq:herschel_bulkely_defn} is formulated as a critical scaling law: $(\Sigma-\Sigma_c)^\beta \sim \dot{\gamma}$, then $\beta = 1/n$ can be viewed as an order parameter exponent, in analogy to static critical phenomena.}

\dkadd{In overdamped systems in general, the flow stress decreases with temperature and increases with driving rate.} Computational studies of thermal activation effects on the yielding transition have been conducted with molecular dynamics simulations of glass formers \cite{chattoraj_universal_2010} and more recently with mesoscale elastoplastic models (EPM) \cite{popovic2021thermalflow,ferrero2021yielding,popovic2021thermalcreep}. \dkadd{While molecular dynamics simulations simulate each atom (or molecule), EPMs coarse grain to the level of elastic blocks. These blocks fluidize once the local stress exceeds a threshold, whereupon they plastically dissipate the local stress. The advantage of this coarse-graining is that one can  simulate for long periods and for large systems, while respecting the underlying symmetries of the yielding transition. } \dkdel{In overdamped systems in general, the flow stress decreases with temperature and increases with driving rate.} Recent EPM studies have mostly focused on the scaling of the flow stress and have proposed that it takes a scaling form (in analogy with the depinning transition \cite{bustingorry_thermal_2007}) 
\begin{equation} 
\dot{\gamma} \sim T^\psi f\left( \frac{\Sigma(T,\dot{\gamma}) - \Sigma_c}{T^{1/\alpha}} \right)
\label{eq:thermal_rounding_expression}
\end{equation}
where $\psi = \beta/\alpha$ is the Fisher thermal rounding exponent, and $\alpha$ is an exponent that characterizes the shape of the energy landscape (see below). 

Relatively little work has focused on the interplay between temperature effects and the scaling description for avalanches. Karmakar et al. also showed that, like with higher driving rates, increased temperatures interfere with avalanches, destroying their anomalous scaling with system size \cite{karmakar_statistical_2010}. In that work, the stress-fluctuation scales can be worked out  by determining the timescales of the system, and identifying when they compete with each other. 

Subsequent to that work, there have been significant theoretical developments in the AQS regime, with scaling theories connecting different scale-free aspects of the yielding transition. One useful advance has been the notion of ``residual stress'' $x = \sigma_{\text{th}} - \sigma$, i.e. the stress necessary to trigger an ST with a threshold $\sigma_{\text{th}}$ in a particular region of the amorphous solid. The distribution of residual stresses  $p(x)\sim x^\theta$ is scale-free in the thermodynamic limit \cite{lin_scaling_2014}, and plays a key role in driving the anomalous stress-fluctuations in these systems. Mesoscale modelling has been particularly helpful in exploring scaling aspects of the yielding transition, as it coarse-grains at the level of STs, and directly exposes $p(x)$\cite{lin_density_2014}.

To that end, in this work we will use a mesoscale model with Arrhenius activation rule (as in refs.~\cite{popovic2021thermalflow,popovic2021thermalcreep,ferrero2021yielding}) 
\begin{equation}
\lambda(x) = \frac{1}{\tau} \exp\left(-\frac{x^\alpha}{T} \right)\,,\label{eq:arrhenius_rate_law}
\end{equation}
to describe temperature and rate effects on the critical behaviour of the yielding transition. This choice of activation rule is motivated by an energy landscape picture: local regions regions of the amorphous solid are stable because there is an energy barrier (scaling as $U\sim C x^\alpha$) preventing their rearrangement (with catastrophe theory suggesting $\alpha =1.5$~\cite{maloney_energy_2006}). Transition state theory suggests that temperature $T$ causes repeated attempts at crossing these barriers, which succeed at a rate proportional to  $\exp(-U(x)/k_B T)$~\cite{truhlar_current_1996}.  In this work, we set $C / k_B = 1$ and the prefactor $\tau^{-1}$ is set so that when $x = 0$ (i.e. no barrier), sites activate on average after $\tau$, thus matching the microscopic timescale for ST rearrangement. \dkadd{We match these two timescales because the exponential prefactor in eq.~\ref{eq:arrhenius_rate_law} is the attempt frequency for barrier crossing. In particle scale simulations of model glasses, this attempt frequency is of order the atomic vibrational frequency, $f_{\text{vib}}\approx \mathcal{O}(1/\tau)$, while the time taken for a plastic rearrangement is of the same order \cite{nicolas_spatiotemporal_2014}.}

With the thermal timescale established, we begin by enumerating the other natural timescales in sheared amorphous solids:
\begin{itemize}
    \item The plastic ST timescale $\tau$ over which atomistic rearrangements occur, i.e. the length of time for which a given ST is fluid and the timescale over which stress is dissipated.   
    \item For a site a distance $x$ from instability, the thermal Arrhenius activation timescale $1/\lambda(x) = \tau \exp(x^\alpha / T)$. In this description, sites with residual stress $x$ have a potential barrier to thermal activation scaling as $U\sim x^\alpha$. 
    \item For a site a distance $x$ from instability, the mechanical yielding timescale $x / \mu\dot{\gamma}$, with $\mu$ a shear modulus.
    \item The average avalanche duration $\langle t \rangle_{\text{av}}$. 
    \item The average loading time between triggering avalanches $\twait$.
\end{itemize}
We  will use the competition of these timescales to sketch out a phase diagram and predict the scaling in different regimes. Then, using numerical simulations, we will confirm the existence of the different phases and verify the scaling laws present in each phase. 

\section{Thermally Activated Elastoplastic Model}
We use a mesoscale elastoplastic model (EPM), which coarse-grains the amorphous solid to a grid of $L^d$ elastically coupled cells. Each cell $i$ has a yield stress $\sigma_{\text{th},i}$ drawn independently from a Weibull distribution (shape parameter $k=2$) \cite{ruscher_residual_2020,liu_elastoplastic_2021} and an initial stress $\sigma_i = 0$. The system is driven at a fixed strain rate $\dot{\gamma}$ which adds stress uniformly to all sites at a global rate $\dot{x} = \mu \dot{\gamma}$. In the following we set $\mu=1$. Sites fluidize immediately when their residual stress $x = \sigma_{\text{th},i} - |\sigma_i| \le 0$. Once fluid, sites remain fluid for $2\tau$, before again becoming elastic. Fluid sites dissipate stress as $\dot{\sigma}_i \sim -\sigma_i / \tau$. We use a finite-element solver to propagate stresses from fluidized sites, which automatically produces the anisotropic Eshelby-like stress-fields characteristic of STs. In contrast to the other thermally activated EPMs~\cite{popovic2021thermalflow,ferrero2021yielding,popovic2021thermalcreep}, we use a real-space stress-propagator more similar to  ~\cite{tyukodi_avalanches_2019,budrikis_universal_2017}. In some of our simulations, we follow refs.~\cite{lin_mean-field_2016,parley_aging_2020} and shuffle the indices of the sites when applying the ``kicks'' from the fluidized sites and refer to these simulations as ``shuffled-kernel'' or mean-field (MF) simulations. There are two features that distinguish our model from the EPMs used in refs.~\cite{ferrero2021yielding,popovic2021thermalflow,popovic2021thermalcreep}. Firstly, we study the system under constant strain-rate, as is more typical in MD simulations. This makes it challenging to numerically sample the time to failure for sites with $x\approx 1$, but can be accelerated for $\alpha = 1,2$, to which we restrict our focus in this work.  Secondly, our EPM implementation does not use periodic boundary conditions. For details on the acceleration algorithm and implementation details, see Appendix~\ref{sec:implementation_details}. 

\section{Competition of timescales}
\begin{figure}
    \centering
    \includegraphics[width=\columnwidth]{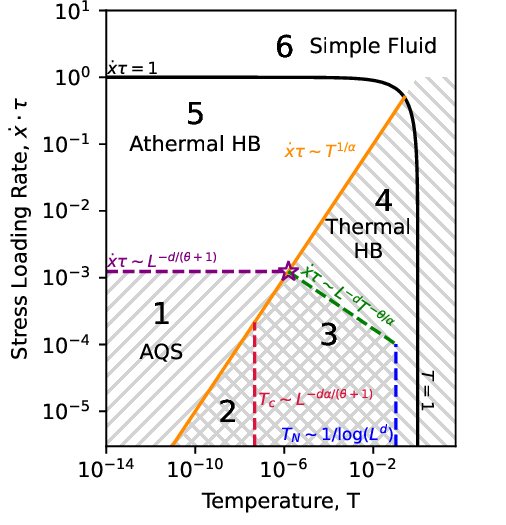}
    \caption{Dynamical phase diagram for the thermal two-dimensional EPM at $L=128$ for $\alpha  = 2$. \dkrep{Dashed}{ Blue, green, red, and purple} lines have finite-size scaling with $L$. ``//'' hatching indicates a region with non-overlapping avalanches. ``\textbackslash\textbackslash'' hatching indicates thermal effects are prevalent.}
    \label{fig:phase_diagram}
\end{figure}

Equation~\ref{eq:arrhenius_rate_law}  implies a characteristic stress scale, \begin{equation}
    x_c = T^{1/\alpha} \,. 
\end{equation}
The Arrhenius activation rate is of order $1/\tau$ when $x \le x_c(T)$. The first two natural timescales to equate are
\begin{equation}
    x_c / \dot{x} = \tau \,,
\end{equation}
which sets the mechanical yielding timescale for a site with residual stress $x_c$ to the thermal yielding time for a site at or below $x_c$. This relation defines the orange line
\begin{equation}\xdotc(T)\equiv\frac{1}{\tau}T^{1/\alpha}\end{equation}
dividing regions 1 and 5 from 2, 3 and 4 in Fig.~\ref{fig:phase_diagram}. \dkadd{For simulations above this orange line with $\dot{x} > \xdotc(T)$}, thermal effects are minimal, since even sites with $x<x_c$ are driven to mechanical failure at $x=0$ before thermal activation can occur. \dkdel{For even higher shearing rates, when $\dot{x}\tau = 1$, sites injected at $x=1$ yield within a single plastic timescale, creating a ``mechanical fluid'' in region 6.}

The division between regions 2 and 3 comes from competition between finite-size scaling and the thermal activation of sites. \dkadd{In both regimes, avalanches are scale-free up to a cut-off scale, set by the either temperature or the linear dimension $L$ of the system.  The size of an avalanche is defined by the stress dissipated by the avalanche, with $S = L^d \delta \Sigma_{\text{av}}$ (approximately the number of yielding sites).}  In steady state, the average stress dissipated by avalanches \dkdel{(whose size $S$ is defined by the stress drop during the avalanche as $S = L^d \delta \Sigma_{\text{av}}$)} must equal the stress loaded between avalanches: $\dot{x} \twait  = \langle \delta \Sigma \rangle_{\text{av}} = L^{-d} \langle S \rangle $. Since $\twait(L)$ scales with $L$, this produces a non-trivial scaling for the mean avalanche size $\langle S\rangle$. In the athermal case, an avalanche begins when the weakest site (left at $x_{min}$ by the preceding avalanche) reaches $x=0$, so in AQS $\twait\dot{x} = \xm$. If $x$ is independent between sites (which appears to be approximately true), $\xm$ is entirely determined by the probability density function for $x$, $p(x)$, with $L^{-d} \sim \int_0^{\xm} p(x') dx'$. However, thermal effects alter the form of $p(x)$ from the power-law form $p(x) \sim x^\theta$ expected in the thermodynamic limit.

\dkadd{Arrhenius activations deplete sites with $x<x_c$ (}as can be seen in Appendix~\ref{sec:px_form}, Fig.~\ref{fig:px_xc_and_xp})\dkdel{, so long as $\dot{x} < x_c / \tau$ (i.e.} in the thermal regime\dkdel{)t sites below $x_c$ are rare}. If we approximate \dkadd{the density of sites with $x<x_c$ as zero} with $p(x) = Ax^\theta$ for $x\in(x_c, 1)$ and elsewhere zero and us\dkadd{ing the relation} $L^{-d} \sim \int_0^{\xm} p(x') dx' $, we obtain:
\begin{equation}
    \xm \sim \left(\frac{\theta+1}{A} L^{-d} + x_c^{\theta+1} \right)^{1/(\theta+1)} \,, \label{eq:xmin_T_activations}
\end{equation}
which is valid whenever thermal activations occur quickly compared to mechanical yielding (i.e. in regions 2, 3, and 4). Clearly, there is a natural temperature scale when $x_c^{\theta+1} \sim L^{-d}$, i.e. when \begin{equation} T_c \sim L^{-d \alpha / (\theta+1)}\,. \label{eq:temperature_cutoff_scaling} \end{equation} This is the red line dividing regions 2 \& 3. The transition across this line is detailed in section~\ref{sec:av_size_cutoff}\dkadd{, but in essence this phase-line captures whether avalanches are truncated by finite-size or by nonzero  temperature.}.

At high driving rates, \dkadd{the time needed for} amorphous solids \dkadd{to plastically dissipate loaded stress becomes longer than the time needed to trigger a new plastic event, which results in} \dkdel{  exhibit } a characteristic shear-thinning stress response, typically captured by the Herschel-Bulkley relation (eq.~\ref{eq:herschel_bulkely_defn}): $\dot{\gamma}^n \sim \Sigma - \Sigma_c$, 
where $n= 1/\beta$ is the Herschel-Bulkley exponent. \dkdel{A typical 2d value for the Herschel-Bulkley exponent found in EPMs is $n \approx 2/3$ \cite{lin_scaling_2014}. In experiments, $n \in [0.4, 1]$ have been reported, and indeed $n$ can vary systematically in response to temperature~\cite{caggioni_variations_2020} and pH~\cite{gutowski_scaling_2012}, though these effects are material specific.}  This rise in stress \dkadd{above the critical flow stress $\Sigma_c$ occurs whenever the stress dissipation timescale (be it from avalanches, individual STs, or a more generic dissipative mechanism) competes with the loading timescale. } \dkdel{with strain rate occurs because} \dkadd{In our case, this first occurs when} avalanches \dkadd{(the longest dissipative timescale in our system)} begin to temporally overlap, i.e. the loading time is comparable to the duration of avalanches \cite{karmakar_statistical_2010}. 

The next several relations determine the onset of avalanche overlap.  
\dkdel{An avalanche continues indefinitely, as long as the time between shear-transformation zone activations is less than $\tau$. Hence, we can establish a maximum shear rate of $\tau\dot{x} = 1$, which divides regions 5 and 6. Similarly, }We can establish a maximum temperature, $T_N$, for which avalanches in a finite-size system remain discrete by setting the rate of activation for $L^d$ freshly injected sites equal to $1/\tau$:
\begin{equation}
    \tau^{-1} =  \frac{L^d}{\tau}e^{-1^\alpha /T_N }
\end{equation}
implying $T_N = 1/\log\left(L^d \right)$. This is the blue line dividing regions 3 and 4 in Fig.~\ref{fig:phase_diagram}.\dkdel{ At an even greater temperature, $T=1$, freshly injected sites at $x=1$ will yield within one plastic timescale $\tau$, setting the threshold for the ``thermal fluid'' in region 7. In this region, the stress accumulated during time $\tau$ is $\Sigma = \tau \cdot \dot{x}$ and so the system is well described by a simple Newtonian fluid (i.e. $\Sigma_c = 0$ and $n = 1$). Since mechanical fluidization can initiate even faster than the Arrhenius rate, region 6 dominates over region 7. }
\dkadd{At lower temperatures, avalanche overlap can be assisted by higher driving rates.} As before, \dkdel{
For non-molten temperatures and lower-rates, }avalanches can  overlap because the loading time between avalanches  is comparable to the plastic time i.e.:
\begin{equation}
    \twait  = \tau\,.
\end{equation}

\begin{table*}[]
    \centering
    \begin{tabular}{|c|c|c|p{50mm}|}\hline 
         \# & Name & Scalings & Notes  \\ \hline \hline
        1 & Athermal quasistatic & $\langle S\rangle_{\text{AQS}} \sim L^{d\theta /(\theta + 1)}$ & Dynamics are as if $T = 0$ and $\dot{\gamma} = 0$\\ \hline 
        2 & Finite-size truncated avalanches &$\langle S\rangle \sim L^{d\theta /(\theta + 1)} + \mathcal{O}(T^{1/\alpha})$ & Minimal temperature effects.  \\ \hline 
        3 & Temperature truncated avalanches &  \begin{tabular}{@{}c@{}} $ \langle S \rangle \sim T^{-\theta /\alpha}$ \\ $\Sigma_{\text{AQS}} - \langle \Sigma\rangle \sim T^{\theta \sigma / (\alpha(2-\tau))}$ \end{tabular}   & Avalanche size is truncated by temperature rather than system size. \\ \hline 
        4 & Thermal Herschel-Bulkley & $\langle \Sigma\rangle = \Sigma_c(T) + \dot{\gamma}^{n_{\text{th}}}$& Avalanches overlap, temperature effects present. \\ \hline 
        5 & Athermal Herschel-Bulkley &$\langle \Sigma\rangle = \Sigma_c + \dot{\gamma}^{n_{\text{AQS}}}$ & Avalanches overlap, no temperature effects\\ \hline 
        6 & Simple Fluid & $\langle \Sigma \rangle \sim \dot{\gamma} $ & Sites yield immediately and independently. \\  \hline 
    \end{tabular}
    \caption{A summary of the derived phases and scaling of avalanche size or stress with temperature or system size. }
    \label{tab:phases_summary}
\end{table*}

We first handle the thermally assisted case (i.e. $\dot{x} < \xdotc(T)$). \dkdel{When}Thermal activation \dkadd{occurs when the weakest site reaches} $x_c$, \dkadd{so the loading time is} $\twait = (\xm - x_c )  / \dot{x}$. For $T > T_c(L)$, we can expand eq.~\ref{eq:xmin_T_activations} to first order in $L^{-d}$ for $\xm$ giving:
\begin{equation}
    \twait = \frac{1}{\dot{x}}\left[ x_c^{-\theta} L^{-d} \right] \,. \label{eq:twait_first_order_small_L}
\end{equation}
Using $\twait = \tau$ and $x_c = T^{1/\alpha}$ we find:
\begin{equation}
    \xdotover (T > T_c(L)) \sim \frac{1}{\tau}L^{-d} T^{-\theta / \alpha}\label{eq:thermal_overlap_threshold}\,.
\end{equation} This gives the green line dividing region 3 and 4 in Fig.~\ref{fig:phase_diagram}. Now for the second case, where \dkrep{$ \dot{x} > \xdotc(T)$ }{$T < (\tau \dot{x})^\alpha$}, we know that thermal activation should be rare (i.e. mechanical effects dominate). Then, the form of $p(x)$ is altered to have a size and drift velocity dependent plateau with \dkadd{$p(x) = p_0$ for $x< x_p(\dot{x},L)$} \dkdel{$  \sim (\tau \dot{x} + 0.02L^{-2})$} (see Appendix~\ref{sec:px_form} Fig.~\ref{fig:px_xc_and_xp} and associated discussion). If we instead use that $\xm$ is in the plateau (an appropriate assumption for large $L$), we have 
\begin{equation}
    \xm = L^{-d} p_0^{-1}
\end{equation}
where $p_0 = A x_p^\theta$. Then, equating the loading time and the ST plastic time, we have:
\begin{equation}
\tau =     \xm / \xdotover \sim \frac{L^{-d}}{\xdotover x_p(\dot{x},L)^\theta} 
\end{equation}
which yields the purple line, dividing region 1 and 5. \dkadd{For large $L$, the onset of the plateau scales as $x_p\sim \tau \dot{x}$, and we have}
\begin{equation}
    \xdotover(T<T_{\text{overlap}}) \sim L^{-d/(1+\theta)}\,.\label{eq:mechanical_overlap_threshold}
\end{equation}
We have found phase lines (eqs.~\ref{eq:thermal_overlap_threshold},\ref{eq:mechanical_overlap_threshold}) for avalanche overlap on both sides of the thermal activation phase line $\xdotc$.  These phase lines should be continuous across the thermal activation phase line, meeting at a system size dependent temperature $T_{\text{overlap}}$:
\begin{equation}
    L^{-d/(1+\theta)} \sim L^{-d}T_{\text{overlap}}^{-\theta / \alpha}
 \implies T_{\text{overlap}}\sim L^{\frac{-d\alpha}{1+\theta}}\label{eq:thermal_overlap_threshold_on_thermal_activity_line}\,,
\end{equation}
which we indicate with the purple star in the phase diagram.\dkadd{ Crucially, $T_{\text{overlap}}(L) \sim T_c(L) \sim L^{-d\alpha/(1+\theta)}$, so that the expansion taken in equation~\ref{eq:twait_first_order_small_L} is always valid. }
Transitions across these lines are explored in section \ref{sec:av_overlap}.

\dkadd{Despite avalanches being ill-defined in the (a)thermal-HB phase, we will show numerically in section-\ref{sec:rheology_transition_over_xcdot} that dynamics are still collective, since the presence of correlations between sites affects the HB exponent $n$. At sufficiently high temperature or driving however, sites yield essentially immediately, destroying any collective effects. When this occurs, the amorphous solid is said to have ``fluidized''. In the fluid phase phase, the stress on any given site $i$  evolves according to $\dot{\sigma}_i = -\sigma_i / \tau + \dot{x}$, which tends to $\sigma_i \rightarrow \tau \dot{x}$. This implies $\langle \Sigma \rangle \sim \dot{\gamma}$ meaning that this phase behaves like a simple Newtonian fluid (hence the name). Fluidization occurs when freshly rejuvenated sites  (initially at $x=1$)   in the absence of noise yield within a time of order the mechanical yielding timescale $\tau$. Since sites with $x<x_c(T)= T^{1/\alpha}$ yield within  about $1\tau$, and it takes time $(1 - x_c)/\dot{x}$ for a site initially at $x=1$ to reach the thermal absorbing barrier, we say that fluidization occurs whenever $\tau\dot{x}>1-T^{1/\alpha}$. This gives the line separating the (a)thermal HB phases (regions 4 and 5) from the fluid phase 6.  }

\section{Numerical tests of the phase diagram}
\begin{figure}
    \centering
    \includegraphics[width=\columnwidth]{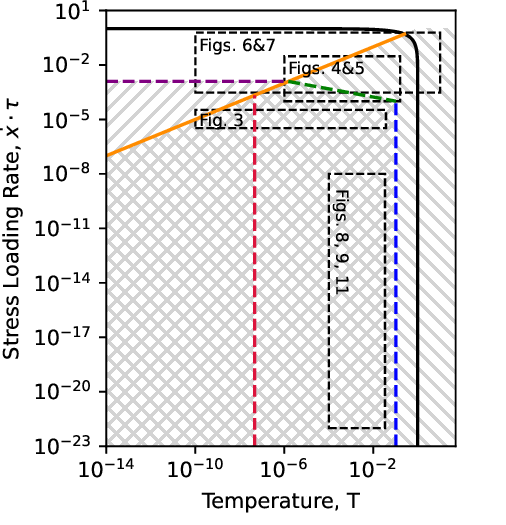}
    \caption{The phase diagram, as in Fig.~\ref{fig:phase_diagram}, but the regions of phase space explored in different figures marked by rectangles. }
    \label{fig:phase_diagram_sim_regions}
\end{figure}

To test our phase diagram and scaling theory, we perform simulations at different stress-loading rates $\dot{x}$, temperatures $T$, system sizes $L$, and with $\alpha = 1$ and $\alpha = 2$, as well as for 2d simulations and 2d simulations with a shuffled kernel. These shuffled kernel simulations keep the broad-tailed kick statistics of the 2d simulations, but remove spatial correlations between sites, thus providing a mean-field (MF) realization of the EPM~\cite{lin_mean-field_2016}. One consequence of this is that the exponent $\theta$ changes from $\theta \approx 0.52$ (in 2d) to $\theta \approx 0.35$, allowing us to vary $\theta$ and test scaling relations involving $\theta$.  For convenience of the reader, Fig.~\ref{fig:phase_diagram_sim_regions} presents an overview of the regions of the phase-diagram that were simulated, and indicates the corresponding figures.

\begin{figure}[t]
    \centering
    \includegraphics[width=\linewidth]{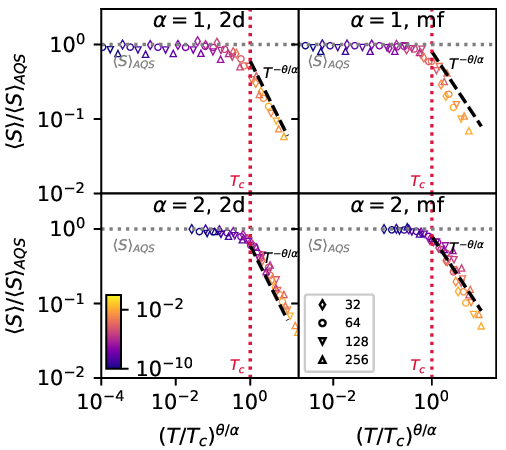}
    \caption{Rescaled mean avalanche sizes for\dkadd{ various temperatures, system sizes, and fixed driving rate} $\dot{x}\tau = 10^{-5}$. Lighter colours are hotter temperatures $T\in (10^{-10},0.037)$,  symbols indicate system sizes ranging from $L=32$ to $L=256$. $\langle S \rangle_{\text{AQS}}$ scales as $\langle S \rangle_{\text{AQS}}\sim L^{d\theta / (\theta+1)}$ and $T_c \sim L^{-d\alpha / (\theta+1)}$\dkadd{, where $\langle S\rangle = \langle S\rangle_{\text{AQS}}$ and $T=T_c$ are indicated with the dotted lines.} }
    \label{fig:size_low_v}
\end{figure}

\subsection{Avalanche size cutoff  \label{sec:av_size_cutoff}}
Discrete avalanches exist in regions 1, 2, and 3. In region 1, avalanches do not overlap and thermal activations do not occur, making this the well-studied AQS limit.

\dkadd{We can study the average avalanche size in regions 1-3 by equating the average stress dissipated in avalanches with the average stress loaded between avalanches. Since avalanche size is defined as $S = L^d \langle \delta \Sigma_{\text{av}}\rangle$, we need merely to compute $\langle \delta \Sigma_{\text{av}}\rangle = \dot{x} \twait$. In zone 3, where thermal effects are large, we expect: $\twait \sim \dfrac{1}{\dot{x}}L^{-d} T^{-\theta/\alpha}$ } (eq.~\ref{eq:twait_first_order_small_L}), so that \begin{equation}\langle S \rangle \sim L^d \dot{x} \twait \sim  T^{-\theta / \alpha}\label{eq:avg_s_T_zone3}\end{equation} (i.e. no system-size dependence of the mean avalanche size). Meanwhile, for $T\ll T_c$ (i.e. $x_c^{1+\theta} \ll L^{-d}$), eq.~\ref{eq:xmin_T_activations} reduces to simply $\xm \sim L^{-d/(\theta+1)}$. For $\dot{x}\twait=\xm - x_c$ with $x_c \ll L^{-d/(1+\theta)}$, we simply have $\langle \delta\Sigma\rangle_{\text{av}} = \dot{x} \twait = \xm$. Hence, in the low-temperature limit, $\langle S\rangle_{\text{AQS}} \sim L^{d\theta / (\theta+1)}$. 

In Fig.~\ref{fig:size_low_v}, we verify this scaling for $\langle S \rangle$ crossing the 1-2 phase line, \dkrep{plotting}{using}  $\langle S\rangle / \langle S\rangle_{\text{AQS}}$ \dkrep{against}{and} $\left( T/T_c\right)^{\theta / \alpha}$, which collapses the AQS plateau and gives a high-temperature tail scaling as $\langle S \rangle \sim \left( (T/T_c)^{\theta/\alpha} \right)^{-1}$. \dkadd{These expressions have no driving rate dependence, and we have verified the crossover and scaling at different driving rates than those show in Fig.~\ref{fig:size_low_v}.} We \dkadd{indeed} find almost no driving-rate dependence in the avalanche size, except for those simulations close to $T_N$ (orange in Fig.~\ref{fig:size_low_v}), where lower velocities weakly decrease avalanche size due to stress-softening. For this reason, the highest temperature fall slightly below the $T^{-\theta/\alpha}$ scaling.

Now, assuming a power-law form for the distribution of avalanche sizes $p(S) \sim S^{-\tau} g(S/S_c(T,L))$, truncated at $S_c(T,L)$ \dkadd{ by finite-system size or temperature}, we have that:
\begin{equation}
    \langle S \rangle = \int_0^\infty S^{1-\tau}g(S/S_c) ds = S_c^{2-\tau} \int_{0}^\infty u^{1-\tau} g(u) du\,,
\end{equation}
implying 
\begin{equation} \langle S \rangle \sim S_c^{2-\tau} \,.\label{eq:avg_s_sc_relation}\end{equation} Coupled with $\langle S\rangle \sim T^{-\theta / \alpha}$ this implies $S_c(T) \sim T^{-\theta/(\alpha(2-\tau))}$. Meanwhile, for $T < T_c(L)$, we have the usual AQS regime scaling, $S_c(L) \sim L^{d/((\theta+1)(2-\tau))}$ and consequently the scaling relation $d_f = d/((\theta+1)(2-\tau))$ \cite{lin_density_2014}. \dkadd{Though this scaling relation} works well for the shuffled kernel simulations, we note that for 2d simulations, there are small additional corrections \dkadd{in the AQS limit} \dkdel{due to finite-size scaling effects in $p(x)$} as described in \cite{korchinski_signatures_2021} that we do not account for here. \dkadd{These additional corrections only affect the slow-driving AQS limit, since they depend on the structure of the plateau in Fig.~\ref{fig:px_xc_and_xp}, which is destroyed by temperature fluctuations. }

Our finding, that there are ``anomalous'' stress fluctuations (i.e. $\langle \delta\Sigma\rangle_{\text{av}} \lesssim L^{-d}=N^{-1}$) below a critical temperature related to the size of the system, and ``normal'' (i.e. $\langle \Delta\Sigma\rangle_{\text{av}} \sim L^{-d} \sim N^{-1}$) fluctuations above this temperature, is consistent with the previous results obtained by Karmakar et al. with particle scale simulations \cite{karmakar_statistical_2010,hentschel_size_2010}. 

\begin{figure}[t]
    \centering
    \includegraphics[width=\linewidth]{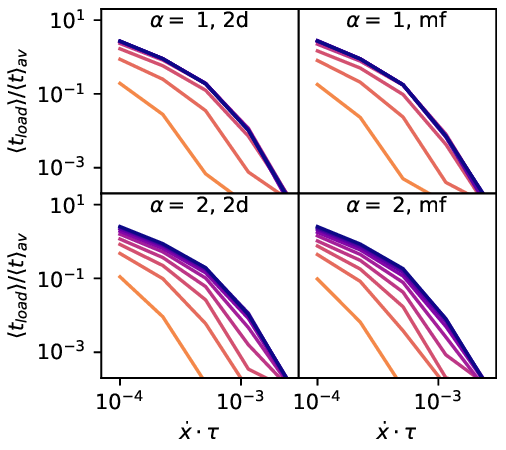}
    \caption{Degree of avalanche overlap, as measured by the ratio of the waiting time and the average avalanche duration, for $L=64$. Lighter colors are hotter temperatures.}
    \label{fig:twait_over_T}
\end{figure}

\subsection{Avalanche overlap onset}
\label{sec:av_overlap}
\dkadd{The onset of Herschel-Bulkley power-law scaling occurs when} the timescale of avalanches is comparable to the timescale between loading, and so stress is added to the system faster than avalanches can release it. \dkadd{What happens when avalanches begin to overlap in time?  As driving rate is increased, spatially distinct and largely non-interacting avalanches are nucleated faster than older avalanches conclude, as was shown by explicitly measuring the correlation length in athermal molecular dynamics~\cite{clemmer_criticality_2021}. Although we do not explicitly measure the correlation length or degree of spatial overlap in our system, we anticipate a similar phenomenon takes hold as temperature is increased, with a regime of spatially distinct but temporally overlapping avalanches. This phase marks the beginning of HB scaling. At sufficiently high strain-rates, the avalanches begin to increasingly overlap and interact, and it is no longer possible to individualize them as independent collective events.  
}

\begin{figure}[t]
    \centering
    \includegraphics[width=\linewidth]{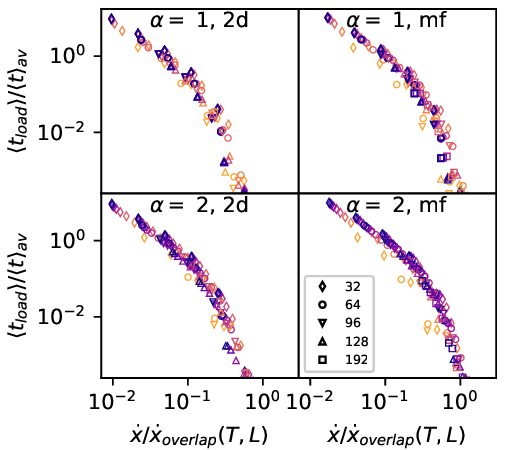}
    \caption{Finite-size scaling collapse for the avalanche overlap parameter, for simulations with different temperatures (lighter colors show higher temperatures) and system sizes indicated by marker shape. }
    \label{fig:fss_twait_collapse}
\end{figure}

\dkadd{Because our system is finite, the HB phase still has brief periods of quiescence with no plastic activity. Since we denote an ``avalanche'' as any period of plasticity bounded by periods of quiescence (of at least $3\tau$), we will be lumping together spatially distinct, but temporally overlapping avalanches. We can detect the onset of this behaviour and consequently test our derived scaling laws for the onset of temporal avalanche overlap, by considering }\dkdel{it is therefore natural to consider} the ratio of time-scales: $\twait / \langle t \rangle_{\text{av}}$. As avalanches \dkadd{temporally} overlap more and more, the durations between avalanches should become exponentially shorter and rarer, as can be seen in Fig.~\ref{fig:twait_over_T}. It is clear that there is a temperature dependence in these curves, but also that for \dkrep{low}{many} temperatures the results are identical. This is consistent with our phase diagram in Fig.~\ref{fig:phase_diagram}, \dkadd{where avalanche overlap starts at fixed $\xdotover(T<T_\text{overlap}(L)) \sim L^{-d/(1+\theta)}$  for low temperatures (eq.~\ref{eq:thermal_overlap_threshold}) and at $\xdotover(T > T_{overlap}(L)) \sim L^{-d}T^{-\theta / \alpha}$ for  high temperatures (eq~\ref{eq:mechanical_overlap_threshold})}. We can capture both scaling behaviours\dkadd{, and their dependence on system size,} with the phenomenological scaling function 
\begin{equation} \tau \dot{x}_{\text{overlap}}(T,L) \sim  L^{-d/(\theta + 1)}(1 + (T/T_{\text{overlap}})^{s \theta /\alpha })^{-1/s} 
\end{equation} which is characterized by a phenomenological sharpness parameter $s$ (which we here take to be 2), and for the overlap temperature $T_{\text{overlap}}  = C L^{-d\alpha /(1+\theta)}$, where $C$ is an arbitrary constant prefactor (eq.~\ref{eq:thermal_overlap_threshold_on_thermal_activity_line}). We find that $C \approx 2$ for $\alpha = 1$ and $C \approx 20$ for $\alpha = 2$ produces an effective collapse for simulations with varying temperature\dkadd{, driving rate,} and system size in Fig.~\ref{fig:fss_twait_collapse}. \dkadd{This confirms our scaling prediction for the onset of avalanche overlap, in both thermal and athermal regimes, and for the finite-size dependence of $T_{\text{overlap}}(L)$.}

\subsection{Rheology transition across $\xdotc(T)$ \label{sec:rheology_transition_over_xcdot}}
\begin{figure}[t]
    \centering
    \includegraphics[width=\linewidth]{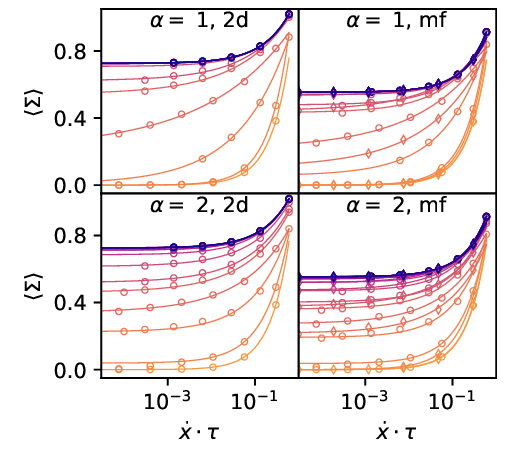}
    \includegraphics[width=\linewidth]{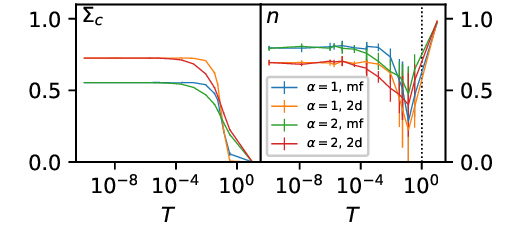}
    \caption{Rheology data with Herschel-Bulkley fits for $L=64$ (circles) and $L=256$ (diamonds),  allowing both exponent $n$ and the temperature dependent stress-plateau $\Sigma_c(T)$ to vary. $T$ varies from $10^{-10}$ to $3\cdot 10^{-1}$. Bottom: fitted $\Sigma_c$ and $n$ values vary with temperature. }
    \label{fig:herschel_bulkley}
\end{figure}

For strain rates above the avalanche-overlap threshold, the Herschel-Bulkley law  $\dot{\gamma} \sim (\langle \Sigma \rangle - \Sigma_c(T))^n$ is a reasonable fit to our data (cf. Fig.~\ref{fig:herschel_bulkley}). 
We  observe that at  high strain-rates \dkdel{($\dot{x} > \xdotc(T)$)}, \dkrep{stress tends towards the athermal value}{stress becomes independent of temperature}. However simulations above a certain temperature show a decrease in the fitted flow stress and in the rheological exponent $n$ (Fig.~\ref{fig:herschel_bulkley}). At still higher temperatures ($T>1$) the rheological exponent tends to $n=1$ as would be expected for a Newtonian fluid. \dkadd{In general, there are minimal system size effects in $\langle \Sigma \rangle$.}

That the rheological exponent changes with $n$ should not surprise experimentalists, where temperature-dependent viscosity effects have been seen to alter the rheological exponent~\cite{caggioni_variations_2020}. However, since the exponent $1/n = \beta$ has been proposed to scale as $\beta = 1 + z/(d-d_f)$ \dkadd{(relating the flow exponent $\beta$ to the dynamical exponent $z$ and fractal dimension $d_f$)} \cite{lin_microscopic_2018}. Naively this would suggest that the avalanche critical exponents would continuously vary with temperature. However, as we show below, this effect can be simply understood as fitting through two exponents operating in different regimes. 

\begin{figure}
    \centering
    \includegraphics[width=\linewidth]{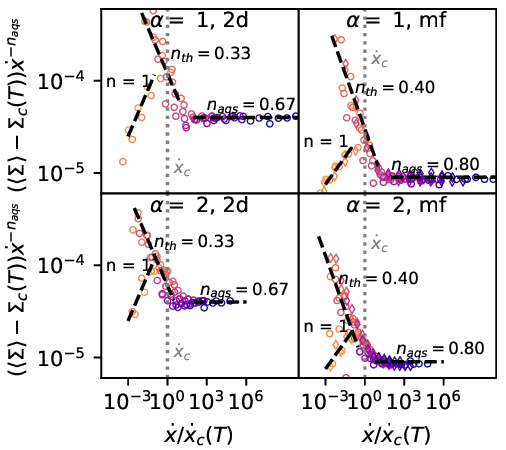}
    \caption{Rheology data, rescaled to exhibit a change in Herschel-Bulkley exponent $n$ for $\dot{x} > \xdotc(T)\sim \frac{1}{\tau}T^{1/\alpha}$, The AQS limit for $n_{\text{AQS}}$ is included in the rescaling to make the transition to the thermal value $n_{\text{th}}$ more evident. Simulations at $T = 10$ in the molten phase (brightest orange) obey $n=1$ scaling. }
    \label{fig:hb_naqs}
\end{figure}

To expose this effect, in Fig.~\ref{fig:hb_naqs} we consider the stress-rise above the flow stress $\Sigma_c(T)$ and divide out the AQS Herschel-Bulkley exponent $n_{\text{AQS}}$. Although the decrease in $n$ appears gradual in Fig.~\ref{fig:herschel_bulkley}, our data in Fig.~\ref{fig:hb_naqs} shows that the transition is actually sharp: there is a thermal and an athermal $n$ exponent. Temperature effects seem to approximately halve the observed Herschel-Bulkley exponent $n$. By studying the stress-rise above $\Sigma_c(T)$, we find that the athermal value $n_{\text{AQS}}$ dominates when $\dot{x} > \xdotc $ (cf. Fig.~\ref{fig:hb_naqs}). The apparent intermediate values of $n$ in Fig.~\ref{fig:herschel_bulkley} are a result of fitting through both regimes. \dkadd{We find the typical 2d AQS value for the Herschel-Bulkley exponent in EPMs ($n \approx 2/3$ \cite{lin_scaling_2014}). We note that in experiments, $n \in [0.4, 1]$ have been reported, and indeed $n$ can vary systematically in response to temperature~\cite{caggioni_variations_2020} or  pH~\cite{gutowski_scaling_2012}.} 

At very high $T$, \dkadd{the system melts into a Newtonian fluid and} the stress scales simply as $\langle \Sigma \rangle \sim \tau \cdot \dot{x}$, i.e. $n =1$. 

\section{Thermally truncated avalanches}
We have provided numerical evidence for the existence of different dynamic phases in our phase diagram. Now we seek to clarify the effect of temperature on the well-defined avalanches in region 3 of the phase diagram. There, temperature is high enough to overcome finite-size effects \dkadd{(in contrast to region 2)}, while remaining low enough that (with sufficiently slow driving) avalanches do not overlap \dkadd{(in contrast to region 4)}. In this case, as we showed previously, the average avalanche size scales as $\langle S \rangle \sim T^{-\theta / \alpha}$, and owing to eq.~\ref{eq:avg_s_sc_relation}, we have that $S_c \sim T^{-\theta/ (\alpha(2-\tau))}$\dkadd{, with minimal system size dependence}. 

However, if driving rates are lowered far below the Herschel-Bulkley onset, a thermal softening of the material occurs as even nominally stable sites above $x_c = T^{1/\alpha}$ can be activated and the mean flow stress $\langle \Sigma\rangle$ is depressed by temperature (see Fig.~\ref{fig:herschel_bulkley}). At extremely low strain-rates (in our data, with $\dot{x}\tau \in (10^{-22}, 10^{-9})$), there is additional softening below the apparent  plateau present in Fig.~\ref{fig:herschel_bulkley}. This softening introduces a stress gap, $\Delta \Sigma (T,\dot{x}) =  \Sigma _{\text{AQS}} - \langle \Sigma \rangle(T,\dot{x})$ (where $\Sigma_{\text{AQS}} \equiv \lim_{\dot{\gamma}\rightarrow 0^+}\langle \Sigma\rangle(T = 0,\dot{\gamma}) = \Sigma_c(T=0)$), which means that, on average, there is less energy available for avalanches to propagate.

\begin{figure}
    \includegraphics[width=\linewidth]{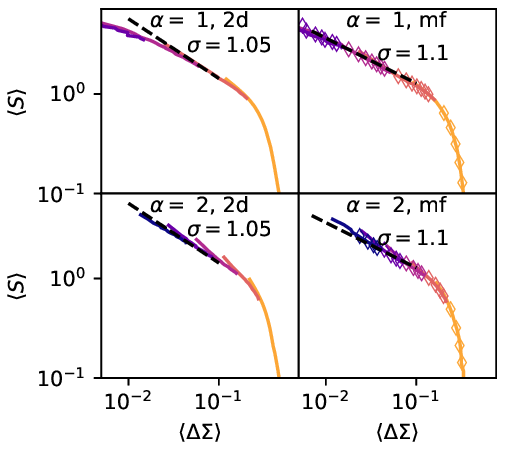}
    \caption{Power-law scaling for mean avalanche size with the stress-gap created at extremely slow driving. Power-law fit uses $\tau = 3/2$ for mean-field (mf) data and $\tau \approx 1.37$ for 2d data. Dashed lines scaling are set by eq.~\ref{eq:avg_s_delta_sigma_relation}, with indicated fit values of $\sigma$. \dkadd{ Lines are $L=64$, with diamonds indicating results for $L=256$}.}
    \label{fig:stress_drop_scaling_slow_driving}
\end{figure}

The stress gap introduces a power-law scaling in avalanche size cutoff, $S_c \sim |\Delta \Sigma|^{-1/\sigma}$, with a new exponent $1/\sigma$.   This exponent has been measured in EPM by fixing the stress of the system with $\Sigma < \Sigma_{\text{AQS}}$ and artificially triggering an avalanche by kicking a random site  \cite{budrikis_universal_2017} or by measuring avalanches in the approach to steady-state flow~\cite{lin_criticality_2015}. However with the temperature dependent stress-gap entering, $1/\sigma$ can probed naturally, by considering the mean-avalanche size for simulations at slow driving, where eq.~\ref{eq:avg_s_sc_relation} implies 
\begin{equation}\langle S \rangle \sim |\Delta \Sigma|^{(\tau-2)/\sigma}\,.\label{eq:avg_s_delta_sigma_relation}\end{equation} 
Our data in Fig.~\ref{fig:stress_drop_scaling_slow_driving} is consistent with $1/\sigma \approx 0.91$\dkadd{, although data is limited to less than a decade for $\alpha = 2$}. To the best of our knowledge, this is the first time this exponent has been measured in a strain-controlled simulation. This value of $\sigma$ is higher than previous reports. In ref.~\cite{budrikis_universal_2017}, where stress was fixed below the flow stress and random sites were kicked, $1/\sigma \approx 0.51$, while in ref.~\cite{lin_criticality_2015}, where avalanches were measured \dkadd{in the transient regime} as the system was loaded to its critical point, $1/\sigma \approx 0.59$  (which we infer from the scaling of eq.~\ref{eq:avg_s_delta_sigma_relation} and their mean avalanche size data). Our explanation for this discrepancy is the following: avalanche propagation depends on the number of sites with small residual stress and on correlations between sites. We propose that \dkadd{the exponent $1/\sigma$} therefore depends on how the stress gap $\Delta\Sigma$ is established. In our case, sites with $x < x_c(T)$ (and for low driving, sites $x\approx x_c$) are suppressed. In the above mentioned works, the $p(x)$ distribution evolved with the stress gap, with substantially lower values of $\theta$ reported. 

\begin{figure}[t]
    \centering
    \includegraphics{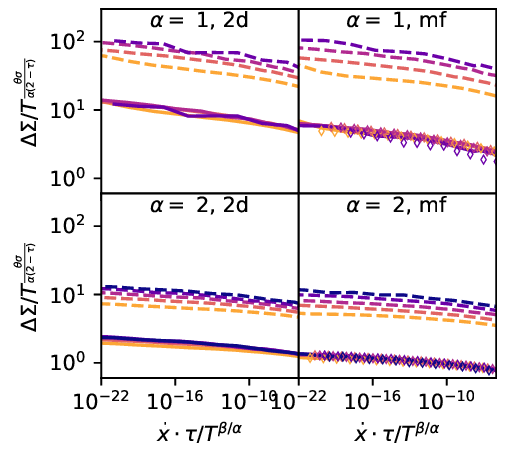}
    \caption{The stress-softening effect, collapsed according to our proposed scaling $\Delta \Sigma \sim T^{\frac{\theta \sigma }{ \alpha(2-\tau)}}$ (solid lines) and according to the previously proposed scaling $\Delta\Sigma \sim T^{1/\alpha}$ (dashed lines) for both for the shuffled kernel and 2d simulations. For clarity, the dashed lines have been shifted vertically by a factor of $3$. All simulations are $L=64$, and $T\in (10^{-4},3\cdot10^{-2})$, see Table.~\ref{tab:exponent_table} for exponent used in collapse.}
    \label{fig:stress_gap_low_strainrate_collapse}
\end{figure}

In any case, if we combine eq.~\ref{eq:avg_s_T_zone3} and eq.~\ref{eq:avg_s_delta_sigma_relation}, we obtain 
\begin{equation}
    \Sigma_{\text{AQS}}   - \langle \Sigma\rangle(T,\dot{\gamma}) = \Delta\Sigma \sim T^{\theta\sigma/(\alpha(2-\tau)) } 
    \label{eq:low_driving_stress_gap}\,,
\end{equation}
which collapses our low strain-rate stress-gap data (cf.  fig.~\ref{fig:stress_gap_low_strainrate_collapse}). This scaling argument does not account for strain rate effects, but we find that these are relatively modest \dkadd{over more than ten decades of strain-rate data}.

In prior work on thermally activated flow in amorphous solids, the characteristic stress scale was identified as $\Delta \Sigma \sim T^{1/\alpha}$,   per eq.~\ref{eq:thermal_rounding_expression} and the strain-rate is scaled by the thermal yielding exponent $\psi=\beta/\alpha$ as  $\dot{\gamma}/T^{\beta/\alpha}$~\cite{ferrero2021yielding,popovic2021thermalflow}. 
We find that this scaling does not effectively collapse our low strain-rate simulations (see the dashed lines in Fig.~\ref{fig:stress_gap_low_strainrate_collapse}). \dkadd{We speculate that the scaling $\Delta \Sigma \sim T^{\theta \sigma /(\alpha(2-\tau))}$ was not previously noted because $\theta \sigma /(\alpha(2-\tau)) \approx 0.8 / \alpha$ is not so dissimilar from $1/\alpha$. The difference in scaling is only obvious when excluding the data from the fast-driving Herschel-Bulkley regime (where the $T^{1/\alpha}$ scaling applies) --- if the data from the fast-driving regime is included, this gap in scaling is visually compressed.  }

\dkadd{Why do different stress scales appear  in slow and fast driving regimes? We propose that in the case of slow-driving, with non-overlapping avalanches, the appropriate stress scale is set by the typical stress dissipated by a single avalanche. This leads to the scaling in eq.~\ref{eq:low_driving_stress_gap}. When avalanches begin to overlap, the pertinent stress scale becomes the stress dissipated by individual STs, for which the relevant stress scale is $x_c\sim T^{1/\alpha}$. Of course, it is also possible that this} alteration in scaling is linked to our use of a strain-controlled driving protocol, while previous simulations used stress-controlled loading.

\dkdel{We do find this $T^{1/\alpha}$ scaling for simulations in the flowing state (i.e.~with overlapping avalanches) in Fig.~\ref{fig:hb_naqs}. 
Meanwhile, with jerky flow  and non-overlapping avalanches, we find a small correction to this scaling, with $\Delta\Sigma \sim T^{0.8/\alpha}$. Why do different stress-scales appear\dkrep{ in slow and fast driving regimes}{  between jerky flow and rheological flow}? In the case of slow driving (non-overlapping avalanches), the appropriate stress-scale is set by the typical stress dissipated by a single avalanche $\langle S\rangle \sim |\Delta\Sigma|^{(\tau-2)/\sigma}$. Once  avalanches overlap the  pertinent stress scale in the problem changes to reflect the stress scale of yielding sites, $x_c\sim T^{1/\alpha}$. }

\dkdel{This correction at low strain-rates may not have been previously noted because the difference in scaling between $T^{1/\alpha}$ and $T^{\approx 0.8/\alpha}$ (see Fig.~\ref{fig:stress_gap_low_strainrate_collapse}) is relatively small. The difference is only obvious when excluding the data from the Herschel Bulkley regime (where the $T^{1/\alpha}$ scaling is correct) --- if the data from the Herschel Bulkley regime is included, this gap is visually compressed.}

\section{Conclusions}
We have derived a schematic dynamic phase diagram for sheared amorphous at finite temperature by comparing the main timescales of the problem. Using EPM simulations, we have provided numerical evidence for these phase lines, by varying the exponent $\alpha$, and by use of a shuffled-kernel, the exponent $\theta$. The main phase lines dictate whether thermal activation occurs and whether avalanches overlap. Avalanche overlap has strong finite-size effects, while the threshold for thermal effects is size independent. In the infinite system size limit, the strict AQS critical point occurs only for $T= 0$ and $\dot{x} = 0$. Avalanches can be truncated by either finite-size effects or by temperature, and we have derived appropriate scaling arguments for both cases. 

Our simulations complement prior stress-controlled work, and in agreement with that work find that  $x_c\sim \Delta\Sigma \sim |\langle \Sigma \rangle - \Sigma_{\text{AQS}}| \sim T^{1/\alpha}$ is the appropriate stress scale when avalanches overlap~\cite{popovic2021thermalflow,ferrero2021yielding}. When avalanches do not overlap, i.e. before Herschel-Bulkley flow onset in the low-strain rate limit, thermal effects soften the material and introduce a new temperature-dependent stress gap. Using scaling arguments, we have linked that stress gap to the stress dissipated by avalanches, finding
$\Delta \Sigma \sim \Sigma_{\text{AQS}} - \langle \Sigma\rangle T^{(\theta\sigma)/(\alpha(2-\tau)}$. Intriguingly, this brings the AQS $1/\sigma$ exponent, which is normally only exposed by stress-controlled simulations, into the thermal problem. Since we arrived at that scaling using generic scaling arguments, it should also hold in stress-controlled simulations.

One avenue that this work does not touch on is aging and thermal history dependence. We use a fixed Weibull distribution for site yield thresholds $\sigma_{\text{th}}$, but \dkrep{the threshold }{this} distribution has \dkdel{been shown to have} temperature \dkadd{and history} dependence~\cite{patinet_connecting_2016,barbot_local_2018}.  To what extent this affects critical behaviour beyond the ductile-brittle transition has not been elucidated.

Another aspect that deserves further study is why the thermal Herschel-Bulkley exponent takes the value it does. Can this be linked by appropriate scaling arguments to the other critical exponents, as is done in the athermal case~\cite{lin_scaling_2014,lin_microscopic_2018}? An accurate space-time exponent $z$ would need to be measured, along with the correlation length exponent $\nu$ (or its thermal equivalent).\dkadd{ We predict a decrease in the correlation length throughout regime 3, where temperature effects truncate avalanche propagation. Although we do not attempt to separate temporally overlapping but spatially distinct avalanches, as might be possible for low driving rates, temperatures and for large systems in the HB regime, we expect correlation lengths will continue to decrease with temperature in the HB regime.}

An interesting observation is that both the thermal and athermal Herschel-Bulkley exponent differ when the kernel is shuffled and correlations between sites are destroyed. \dkadd{This highlights that, although the correlation length in the thermal regime is likely shorter ranged, collective events are still playing a role. With shuffled kernel simulations, we notice that the flow stress is much lower than in full 2d simulations. This indicates that systems with correlated noise evolve to rather different steady states. Does this in turn produce different types of mechanical noise, beyond just that expected by having line-like plastic events? If so, is it enough to simply plug in the different noise distributions at the mean-field level (in the spirit of \cite{lin_mean-field_2016,ferrero_properties_2021}), to explain the observed Herschel-Bulkley exponent for the shuffled kernel? In another sense of ``mean-field'', however, it seems like the correlations that can build up between sites in low dimension matter. Since dimension seems to matter, then we are presumably below the upper critical dimension, which suggests } \dkdel{This seems to suggest} that not all aspects of the yielding transition are mean-field.

\begin{acknowledgments}
This research was undertaken thanks, in part, to funding from the Canada First Research Excellence Fund, Quantum Materials and Future Technologies Program. DK thanks NSERC for financial support through a CGS-D scholarship.
\end{acknowledgments}

\appendix 
\noindent 

\section{Implementation Details\label{sec:implementation_details}}
\subsection{Loading times}
Sites liquefy by thermal activations at rate $\lambda(x)$, or immediately at $x=0$. To simulate the dynamics of our systems, we essentially need to work out which site $i$ will fail next and at what time $\delta t$ (while correctly accounting for loaded stress $\dot{x}$). Once the next site to liquefy is known, stresses propagated from that site can be worked out using the finite element solver, and stress can be relaxed at the failing site. \dkadd{During the time increment $\delta t$,} all sites have their stresses increased by $\dot{x}\cdot \delta t$ and any propagated stresses or relaxations are \dkadd{scaled by a factor of} $\sim \exp[-\delta t/\tau]$\dkdel{ applied}.

Working out the inter-event period $\delta t$ and site $i$ is relatively trivial for fixed stress simulations between avalanches, where activations happen at fixed rate $\lambda(x)$. Since activations are independent Poisson processes, with an exponentially decaying waiting time probability distribution function $p(t) = \lambda(x_i)e^{-\lambda(x_i)t}$, for a system with $N = L^d$ sites one could sample $N$ random numbers $\{R_i\in[0,1)\}$, and find the next activation time for each site by inverting the cumulative distribution function for the exponentially distributed waiting times, so that each site is assigned a time: 
$t_i =\frac{-1}{\lambda(x_i)}\log(1-R_i)$. By finding the $i$ with the smallest $t_i$, one has found the first site the yield, and the appropriate interval $\delta t = t_i$. This naive scheme could obviously be improved by using a global rate, $\lambda = \sum_i \lambda(x_i)$ and choosing a site $i$ at random with weight $\lambda(x_i)$, which reduces the problem to requiring only two random numbers, instead of $N$.

In our system, because $\sigma_i$ evolves in time \dkadd{between ST events} as
\begin{equation}
    \sigma_i(t) = \sigma_{i,0} + \dot{x}t + \sigma^{\text{nonlocal}}_i(1- e^{-t/\tau})
\end{equation}
where $\sigma_i^{\text{nonlocal}}$ are the stresses propagating from liquefied sites to site $i$, the Poisson rates $\lambda(x_i)$ are not constant. We follow the ``naive'' approach, but generalized to inhomogeneous Poisson processes, for each site sampling a random number $R_i \in [0,1)$, and solving for $t_i$ as:  
\begin{equation}
    R_i = P(t<t_i) = \exp\left[-\int_0^{t_i} \lambda(x_i(t)) \dup t\right] \,. \label{eq:R_i_integral_equation}
\end{equation}
As before, once we have a $t_i$ for each site, we pick the lowest $t_i$ for $\delta t$ and liquefy site $i$. In practice, we solve eq.~\ref{eq:R_i_integral_equation} in two ways: (i) after all $\sigma_i^{\text{nonlocal}}$ have decayed by 25$\tau$ and are negligible, $\sigma_i(t)$ are linear in time, and eq.~\ref{eq:R_i_integral_equation} can be solved and inverted analytically for $\alpha = 1$ and $\alpha=2$ (ii) when $\sigma_i(t)$ are non-linear, we solve eq.~\ref{eq:R_i_integral_equation} numerically by constructing and solving a related initial value problem (IVP). 

For the analytical case, using $\lambda(x) = \frac{1}{\tau}\exp(-x^\alpha / T)$ we restrict ourselves to $\alpha = 1$ and $\alpha =2$. For $\alpha = 1$, with $x(t) = \sigma_{\text{th}} - (\sigma_{0} - \dot{x}t)$ (suppressing the ${}_i$ subscripts for brevity), we have
\begin{equation*}
\int_0^{t_i}\lambda(x(t))\dup t = \frac{T}{\dot{x}\tau}e^{-(\sigma_{\text{th}}-\sigma_0)/T}\left(e^{t_i\dot{x}/T}-1 \right)\,.
\end{equation*}
Since $R_i = 1-e^{\int_0^{t_i} \lambda(x_i(t))\dup t}$, we can solve for $t_i$ as:
\begin{align*}
    t_i = \frac{\sigma_{\text{th}}-\sigma_0}{\dot{x}} + \frac{T}{\dot{x}}\log\left[e^{-(\sigma_{\text{th}} - \sigma_0)/T}+\log(1-R_i)\frac{\tau\dot{x}}{T}\right]\,,
\end{align*}
where we can recognize the first term as the mechanical yielding timescale $(\sigma_{\text{th}}-\sigma_0)/\dot{x} = x/\mu\dot{\gamma}$, and the second as a temperature-dependent stochastic correction that can reduce the time to yield. The $\alpha =2$ case is similar, and gives:
\begin{align*}
t_i = &+\frac{\sigma_{\text{th}} - \sigma_0}{\dot{x}}\\ &-\frac{\sqrt{T}}{\dot{x}}{\erf}^{-1}\left[\erf\left(\frac{\sigma_{\text{th}}-\sigma_0}{\sqrt{T}}\right) +\frac{2\dot{x}\tau}{\sqrt{\pi T}}\log(1-R_i) \right]  \,.
\end{align*}
These analytical equations are used between avalanches for potentially very long loading periods. This avoids a potentially expensive numerical integration at each site, since a small time-step (comparable to $\tau$) is necessary to avoid missing the thermal activation of a site (since a site approaching $x\approx x_c = T^{1/\alpha}$ activates on average on a timescale $\approx \tau$). In practice, since $x \equiv \sigma_{\text{th}} - |\sigma|$, we also consider the case $x(t) = \sigma_{\text{th}} + \sigma_0 - \dot{x}t$ to catch the (rare) events in which negatively stressed sites yield thermally. Additionally, these equations are prone to numerical under/overflows, so care must be taken when implementing these equations to deal with potential numerical pitfalls. 

The numerical description is conceptually simpler, though more computationally expensive. To solve eq.~\ref{eq:R_i_integral_equation} numerically, consider the obviously related function $R(t) = \exp\left[ -\int_0^t\lambda(x_i(t'))\dup t'\right]$, for which $R(t_i) = R_i$ is our desired solution. $R(t)$ obeys the following differential equation: 
\begin{equation}
    \ld{R}{t} = (1-R(t)) \lambda(x_i(t))\,,
\end{equation}
for which the IVP $R(t=0)=0$ and $R(t_i)=R_i$ (where $R_i$ is still randomly drawn from $[0,1)$) has a unique solution. We use SciPy's \verb|solve_ivp| routine~ \cite{2020SciPy-NMeth} to integrate these equations for all sites simultaneously -- halting when either $R(t_i) = R_i$ for a site or $x_i =0$ for a site. As in the analytical case, we consider both cases of $x_i = \sigma_{\text{th}} \pm \sigma_i(t)$, allowing for a site to yield because it is stressed too far in either direction. Since the Eshelby-like  stress-propagator has both positive and negative kicks, sites frequently fail in either direction during an avalanche (though they are biased to fail in the forward loading direction).

\subsection{Stress Propagator and Shuffled Kernel}
When sites liquefy, they reduce their stress and redistribute stresses elsewhere in the system. We use the finite-element method on a regular triangular mesh to determine the stress propagation between sites. Each square site in the system consists of four finite-element triangles terminating in a central vertex. We use first-order Lagrange elements for the displacement field $\vc{u}$ \dkadd{(with simple-shear fixed boundary conditions $\vc{u}((x,y)\in \partial\Omega) = (\gamma \cdot y,0)^T$)}, and zeroth-order discontinuous Galerkin elements to represent the stresses, strains, and plastic strains on the plaquettes. We relate the total strain to the displacement as, \begin{equation}
    \tens{\gamma} = \frac{1}{2}\left[ (\vc{\nabla}\,\vc{u})+(\vc{\nabla}\,\vc{u})^T \right]
\end{equation}
and decompose the strain tensor into plastic (stress-free) and elastic parts as: $\tens{\gamma} = \tens{\gamma_{\text{pl}}} + \tens{\gamma_{\text{el}}}$. The elastic-strain contributes to the tensorial stress as:
\begin{equation} 
\tens{\sigma} = 2 \mu  \tens{\gamma_{\text{el}}} + \lambda \tr (\tens{\gamma_{\text{el}}})  \mathds{1} \,,
\end{equation}
where, for our simulations, we use $\mu = 1$ and $\lambda = \frac{1}{2}$. The local stress $\sigma_i$ at a site $i$ is the average $\tens{\sigma}_{xy}$ component of the stress over the square cell. To work out the long-time stress-field from a liquefied site, we increment the plastic strain at that site (equally distributed over all triangles) by the stress at the site. We then solve the elastic equations, and find the stress-increments $\delta \sigma_{ij}$ at all sites in the system. For each site $j$ in the system, we then set $\sigma^{\text{nonlocal}}_j \rightarrow \sigma^{\text{nonlocal}}_j + \delta\sigma_{ij}$, so that this stress-increment is applied exponentially over the next several time units $\tau$.

For the shuffled-kernel simulations, we first identify a central site $i$ in the system (e.g. for $L =32$ at $(x,y) = (15,15)$). We then apply a nominal plastic strain of magnitude $1$, and work out the resulting $\delta\sigma_{ij}$ at all sites. The resulting set of $N-1$ stress-increments are $\{\delta\sigma_{ij}\}$ then stored for later use, so that the expensive finite-element calculation is not repeated throughout the mean-field simulations. When a site $k$ in the shuffled-kernel simulation fails, all sites receive a stress-increment $\sigma_j^{\text{nonlocal}} \rightarrow \sigma_j^{\text{nonlocal}} + \sigma_k \delta\sigma_{ij'}$, where the  $j'$ are drawn without replacement from $\{1, 2,\ldots i-1, i+1, \ldots N\}$. The failing site of course receives the stress-increment $\sigma_k^{\text{nonlocal}} = \sigma_{k}\delta\sigma_{ii}\approx -\sigma_k$. In this way, the stress increments initially calculated for a central site are shuffled for the non-failing site, and scaled according to the stress at the failing site, and the failing site relaxes in accordance to its stress. 

\section{$p(x)$ form\label{sec:px_form}}
\begin{figure}
    \includegraphics[width=\linewidth]{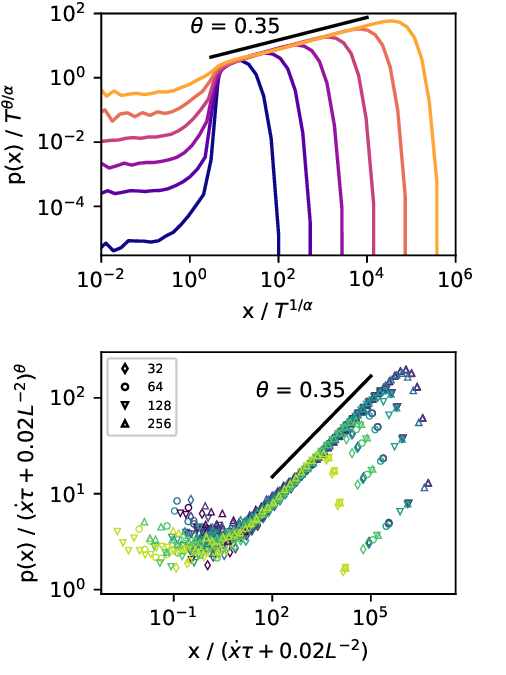}
    \caption{The residual stress distributions of shuffled kernel simulations ($\theta = 0.35$), rescaled by characteristic scales set by temperature, velocity, or finite-size effects. Top: $p(x)$ for simulations with different values of $T$ (brighter colours have higher T) and $L=256$, rescaled by $x_c = T^{1/\alpha}$ for low velocities. Bottom: $p(x)$ for simulations with different values of $v$ (brighter colours have higher strain rate) and $L$ (indicated by symbol), rescaled by $x_p = (\tau \dot{x} + 0.02L^{-2})$.}
    \label{fig:px_xc_and_xp}
\end{figure}

For simulations with well-defined avalanches, $p(x)$ can be sampled immediately after each avalanche. The distribution of $p(x)$ then can be used to predict the loading time before the next avalanche. For simulations in with $\dot{x} < \xdotc(T)$, we find that $p(x)$ takes the form of a power-law with an exponential cut-off at $x < x_c = T^{1/\alpha}$, as can be seen in Fig.~\ref{fig:px_xc_and_xp}. For simulations with a higher-driving rate however, there is a system-size dependent and velocity dependent plateau (Fig.~\ref{fig:px_xc_and_xp}). We expect the velocity dependent plateau to occur at $x \approx \tau \cdot \dot{x}$, while the system-size dependent terminal plateau should scale as $x\sim L^{-d}$~\cite{tyukodi_avalanches_2019,korchinski_signatures_2021}. We approximate  the plateau as occurring at $x_p = (\tau \dot{x} + 0.02L^{-2})$, which effectively collapses the plateau onset in Fig.~\ref{fig:px_xc_and_xp}.

\section{Exponents and scaling relations}
When avalanches are well defined, the maximum avalanche size $S_c$ is set by either temperature effects or finite size effects. In the  case of finite size effects, the fractal dimension of avalanches enters, with $S_c\sim L^{d_f}$. Our scaling description based on a simple truncated power-law for $p(x)$ finds that $d_f = d/((\theta+1)(2-\tau)$ in the athermal limit. When temperature effects are larger than finite-size effects, the maximum avalanche size scales as $S_c(T)\sim T^{-\theta / (\alpha(2-\tau))}$ or as $S_c \sim (\Delta \Sigma)^{-1/\sigma}$. 

\begin{table}[]
    \centering
    \begin{tabular}{c|c|c|c}
        Exponent & Definition & Value (2d) & Value (mf)  \\ \hline \hline 
         $\tau$ & $p(S)\sim s^{-\tau} G(S/S_c)$  & 1.37$\pm0.07$ & 1.5$\pm0.05$ \\ \hline 
         $\sigma$ & $S_c\sim (\Delta\Sigma)^{-1/\sigma}$ & 1.05$\pm0.08$ & 1.1$\pm 0.06$ \\ \hline 
         $\theta$ & $p(x) \sim x^\theta$ & 0.52$\pm0.05$ & 0.35$\pm0.03$\\ \hline 
         $n$ (athermal) & $\Sigma = \Sigma_c + C \dot{\gamma}^n$ & 0.67 $\pm0.02$& 0.8$\pm 0.02$ \\ \hline 
         $n$ (thermal & $\Sigma = \Sigma_c(T) + C \dot{\gamma}^n$ & 0.33$\pm 0.06$ & 0.4 $\pm0.04$ \\ \hline 
         $n$ (molten) & $\Sigma = \tau\mu \dot{\gamma}^n$ & $1\pm 0.01$ & $1\pm 0.01$ 
    \end{tabular}
    \caption{Numerically observed exponents for the thermal EPM model, with $\pm$ errors representing acceptable fit ranges. }
    \label{tab:exponent_table}
\end{table}


\clearpage
\bibliography{biblio}

\begin{thebibliography}{54}%
\makeatletter
\providecommand \@ifxundefined [1]{%
 \@ifx{#1\undefined}
}%
\providecommand \@ifnum [1]{%
 \ifnum #1\expandafter \@firstoftwo
 \else \expandafter \@secondoftwo
 \fi
}%
\providecommand \@ifx [1]{%
 \ifx #1\expandafter \@firstoftwo
 \else \expandafter \@secondoftwo
 \fi
}%
\providecommand \natexlab [1]{#1}%
\providecommand \enquote  [1]{``#1''}%
\providecommand \bibnamefont  [1]{#1}%
\providecommand \bibfnamefont [1]{#1}%
\providecommand \citenamefont [1]{#1}%
\providecommand \href@noop [0]{\@secondoftwo}%
\providecommand \href [0]{\begingroup \@sanitize@url \@href}%
\providecommand \@href[1]{\@@startlink{#1}\@@href}%
\providecommand \@@href[1]{\endgroup#1\@@endlink}%
\providecommand \@sanitize@url [0]{\catcode `\\12\catcode `\$12\catcode
  `\&12\catcode `\#12\catcode `\^12\catcode `\_12\catcode `\%12\relax}%
\providecommand \@@startlink[1]{}%
\providecommand \@@endlink[0]{}%
\providecommand \url  [0]{\begingroup\@sanitize@url \@url }%
\providecommand \@url [1]{\endgroup\@href {#1}{\urlprefix }}%
\providecommand \urlprefix  [0]{URL }%
\providecommand \Eprint [0]{\href }%
\providecommand \doibase [0]{https://doi.org/}%
\providecommand \selectlanguage [0]{\@gobble}%
\providecommand \bibinfo  [0]{\@secondoftwo}%
\providecommand \bibfield  [0]{\@secondoftwo}%
\providecommand \translation [1]{[#1]}%
\providecommand \BibitemOpen [0]{}%
\providecommand \bibitemStop [0]{}%
\providecommand \bibitemNoStop [0]{.\EOS\space}%
\providecommand \EOS [0]{\spacefactor3000\relax}%
\providecommand \BibitemShut  [1]{\csname bibitem#1\endcsname}%
\let\auto@bib@innerbib\@empty
\bibitem [{\citenamefont {Bonn}\ \emph {et~al.}(2017)\citenamefont {Bonn},
  \citenamefont {Denn}, \citenamefont {Berthier}, \citenamefont {Divoux},\ and\
  \citenamefont {Manneville}}]{bonn_yield_2017}%
  \BibitemOpen
  \bibfield  {author} {\bibinfo {author} {\bibfnamefont {D.}~\bibnamefont
  {Bonn}}, \bibinfo {author} {\bibfnamefont {M.~M.}\ \bibnamefont {Denn}},
  \bibinfo {author} {\bibfnamefont {L.}~\bibnamefont {Berthier}}, \bibinfo
  {author} {\bibfnamefont {T.}~\bibnamefont {Divoux}},\ and\ \bibinfo {author}
  {\bibfnamefont {S.}~\bibnamefont {Manneville}},\ }\href
  {https://doi.org/10.1103/RevModPhys.89.035005} {\bibfield  {journal}
  {\bibinfo  {journal} {Reviews of Modern Physics}\ }\textbf {\bibinfo {volume}
  {89}},\ \bibinfo {pages} {035005} (\bibinfo {year} {2017})},\ \bibinfo {note}
  {publisher: American Physical Society}\BibitemShut {NoStop}%
\bibitem [{\citenamefont {Schuh}\ \emph {et~al.}(2007)\citenamefont {Schuh},
  \citenamefont {Hufnagel},\ and\ \citenamefont
  {Ramamurty}}]{schuh_mechanical_2007}%
  \BibitemOpen
  \bibfield  {author} {\bibinfo {author} {\bibfnamefont {C.~A.}\ \bibnamefont
  {Schuh}}, \bibinfo {author} {\bibfnamefont {T.~C.}\ \bibnamefont
  {Hufnagel}},\ and\ \bibinfo {author} {\bibfnamefont {U.}~\bibnamefont
  {Ramamurty}},\ }\href {https://doi.org/10.1016/j.actamat.2007.01.052}
  {\bibfield  {journal} {\bibinfo  {journal} {Acta Materialia}\ }\textbf
  {\bibinfo {volume} {55}},\ \bibinfo {pages} {4067} (\bibinfo {year}
  {2007})}\BibitemShut {NoStop}%
\bibitem [{\citenamefont {Manning}\ \emph {et~al.}(2009)\citenamefont
  {Manning}, \citenamefont {Daub}, \citenamefont {Langer},\ and\ \citenamefont
  {Carlson}}]{manning_rate-dependent_2009}%
  \BibitemOpen
  \bibfield  {author} {\bibinfo {author} {\bibfnamefont {M.~L.}\ \bibnamefont
  {Manning}}, \bibinfo {author} {\bibfnamefont {E.~G.}\ \bibnamefont {Daub}},
  \bibinfo {author} {\bibfnamefont {J.~S.}\ \bibnamefont {Langer}},\ and\
  \bibinfo {author} {\bibfnamefont {J.~M.}\ \bibnamefont {Carlson}},\ }\href
  {https://doi.org/10.1103/PhysRevE.79.016110} {\bibfield  {journal} {\bibinfo
  {journal} {Physical Review E}\ }\textbf {\bibinfo {volume} {79}},\ \bibinfo
  {pages} {016110} (\bibinfo {year} {2009})},\ \bibinfo {note} {publisher:
  American Physical Society}\BibitemShut {NoStop}%
\bibitem [{\citenamefont {Fielding}(2014)}]{fielding_shear_2014}%
  \BibitemOpen
  \bibfield  {author} {\bibinfo {author} {\bibfnamefont {S.~M.}\ \bibnamefont
  {Fielding}},\ }\href {https://doi.org/10.1088/0034-4885/77/10/102601}
  {\bibfield  {journal} {\bibinfo  {journal} {Reports on Progress in Physics}\
  }\textbf {\bibinfo {volume} {77}},\ \bibinfo {pages} {102601} (\bibinfo
  {year} {2014})},\ \bibinfo {note} {publisher: IOP Publishing}\BibitemShut
  {NoStop}%
\bibitem [{\citenamefont {Barlow}\ \emph {et~al.}(2020)\citenamefont {Barlow},
  \citenamefont {Cochran},\ and\ \citenamefont
  {Fielding}}]{barlow_ductile_2020}%
  \BibitemOpen
  \bibfield  {author} {\bibinfo {author} {\bibfnamefont {H.~J.}\ \bibnamefont
  {Barlow}}, \bibinfo {author} {\bibfnamefont {J.~O.}\ \bibnamefont
  {Cochran}},\ and\ \bibinfo {author} {\bibfnamefont {S.~M.}\ \bibnamefont
  {Fielding}},\ }\href {https://doi.org/10.1103/PhysRevLett.125.168003}
  {\bibfield  {journal} {\bibinfo  {journal} {Physical Review Letters}\
  }\textbf {\bibinfo {volume} {125}},\ \bibinfo {pages} {168003} (\bibinfo
  {year} {2020})},\ \bibinfo {note} {publisher: American Physical
  Society}\BibitemShut {NoStop}%
\bibitem [{\citenamefont {Argon}(1979)}]{argon_plastic_1979}%
  \BibitemOpen
  \bibfield  {author} {\bibinfo {author} {\bibfnamefont {A.~S.}\ \bibnamefont
  {Argon}},\ }\href {https://doi.org/10.1016/0001-6160(79)90055-5} {\bibfield
  {journal} {\bibinfo  {journal} {Acta Metallurgica}\ }\textbf {\bibinfo
  {volume} {27}},\ \bibinfo {pages} {47} (\bibinfo {year} {1979})}\BibitemShut
  {NoStop}%
\bibitem [{\citenamefont {Argon}\ and\ \citenamefont
  {Kuo}(1979)}]{argon_plastic_1979-bubble}%
  \BibitemOpen
  \bibfield  {author} {\bibinfo {author} {\bibfnamefont {A.~S.}\ \bibnamefont
  {Argon}}\ and\ \bibinfo {author} {\bibfnamefont {H.~Y.}\ \bibnamefont
  {Kuo}},\ }\href@noop {} {\bibfield  {journal} {\bibinfo  {journal} {Materials
  science and Engineering}\ }\textbf {\bibinfo {volume} {39}},\ \bibinfo
  {pages} {101} (\bibinfo {year} {1979})}\BibitemShut {NoStop}%
\bibitem [{\citenamefont {Maloney}\ and\ \citenamefont
  {Lemaître}(2006)}]{maloney_amorphous_2006}%
  \BibitemOpen
  \bibfield  {author} {\bibinfo {author} {\bibfnamefont {C.~E.}\ \bibnamefont
  {Maloney}}\ and\ \bibinfo {author} {\bibfnamefont {A.}~\bibnamefont
  {Lemaître}},\ }\href {https://doi.org/10.1103/PhysRevE.74.016118} {\bibfield
   {journal} {\bibinfo  {journal} {Physical Review E}\ }\textbf {\bibinfo
  {volume} {74}},\ \bibinfo {pages} {016118} (\bibinfo {year} {2006})},\
  \bibinfo {note} {publisher: American Physical Society}\BibitemShut {NoStop}%
\bibitem [{\citenamefont {Nicolas}\ \emph {et~al.}(2015)\citenamefont
  {Nicolas}, \citenamefont {Puosi}, \citenamefont {Mizuno},\ and\ \citenamefont
  {Barrat}}]{nicolas_elastic_2015}%
  \BibitemOpen
  \bibfield  {author} {\bibinfo {author} {\bibfnamefont {A.}~\bibnamefont
  {Nicolas}}, \bibinfo {author} {\bibfnamefont {F.}~\bibnamefont {Puosi}},
  \bibinfo {author} {\bibfnamefont {H.}~\bibnamefont {Mizuno}},\ and\ \bibinfo
  {author} {\bibfnamefont {J.-L.}\ \bibnamefont {Barrat}},\ }\href@noop {}
  {\bibfield  {journal} {\bibinfo  {journal} {Journal of the Mechanics and
  Physics of Solids}\ }\textbf {\bibinfo {volume} {78}},\ \bibinfo {pages}
  {333} (\bibinfo {year} {2015})}\BibitemShut {NoStop}%
\bibitem [{\citenamefont {Albaret}\ \emph {et~al.}(2016)\citenamefont
  {Albaret}, \citenamefont {Tanguy}, \citenamefont {Boioli},\ and\
  \citenamefont {Rodney}}]{albaret2016mapping}%
  \BibitemOpen
  \bibfield  {author} {\bibinfo {author} {\bibfnamefont {T.}~\bibnamefont
  {Albaret}}, \bibinfo {author} {\bibfnamefont {A.}~\bibnamefont {Tanguy}},
  \bibinfo {author} {\bibfnamefont {F.}~\bibnamefont {Boioli}},\ and\ \bibinfo
  {author} {\bibfnamefont {D.}~\bibnamefont {Rodney}},\ }\href@noop {}
  {\bibfield  {journal} {\bibinfo  {journal} {Physical Review E}\ }\textbf
  {\bibinfo {volume} {93}},\ \bibinfo {pages} {053002} (\bibinfo {year}
  {2016})}\BibitemShut {NoStop}%
\bibitem [{\citenamefont {Nicolas}\ and\ \citenamefont
  {Rottler}(2018)}]{nicolas_orientation_2018}%
  \BibitemOpen
  \bibfield  {author} {\bibinfo {author} {\bibfnamefont {A.}~\bibnamefont
  {Nicolas}}\ and\ \bibinfo {author} {\bibfnamefont {J.}~\bibnamefont
  {Rottler}},\ }\href {https://doi.org/10.1103/PhysRevE.97.063002} {\bibfield
  {journal} {\bibinfo  {journal} {Physical Review E}\ }\textbf {\bibinfo
  {volume} {97}},\ \bibinfo {pages} {063002} (\bibinfo {year}
  {2018})}\BibitemShut {NoStop}%
\bibitem [{\citenamefont {Lin}\ \emph {et~al.}(2014{\natexlab{a}})\citenamefont
  {Lin}, \citenamefont {Lerner}, \citenamefont {Rosso},\ and\ \citenamefont
  {Wyart}}]{lin_scaling_2014}%
  \BibitemOpen
  \bibfield  {author} {\bibinfo {author} {\bibfnamefont {J.}~\bibnamefont
  {Lin}}, \bibinfo {author} {\bibfnamefont {E.}~\bibnamefont {Lerner}},
  \bibinfo {author} {\bibfnamefont {A.}~\bibnamefont {Rosso}},\ and\ \bibinfo
  {author} {\bibfnamefont {M.}~\bibnamefont {Wyart}},\ }\href@noop {}
  {\bibfield  {journal} {\bibinfo  {journal} {Proceedings of the National
  Academy of Sciences}\ }\textbf {\bibinfo {volume} {111}},\ \bibinfo {pages}
  {14382} (\bibinfo {year} {2014}{\natexlab{a}})}\BibitemShut {NoStop}%
\bibitem [{\citenamefont {Sandfeld}\ \emph {et~al.}(2015)\citenamefont
  {Sandfeld}, \citenamefont {Budrikis}, \citenamefont {Zapperi},\ and\
  \citenamefont {Castellanos}}]{sandfeld_avalanches_2015}%
  \BibitemOpen
  \bibfield  {author} {\bibinfo {author} {\bibfnamefont {S.}~\bibnamefont
  {Sandfeld}}, \bibinfo {author} {\bibfnamefont {Z.}~\bibnamefont {Budrikis}},
  \bibinfo {author} {\bibfnamefont {S.}~\bibnamefont {Zapperi}},\ and\ \bibinfo
  {author} {\bibfnamefont {D.~F.}\ \bibnamefont {Castellanos}},\ }\href@noop {}
  {\bibfield  {journal} {\bibinfo  {journal} {Journal of Statistical Mechanics:
  Theory and Experiment}\ }\textbf {\bibinfo {volume} {2015}},\ \bibinfo
  {pages} {P02011} (\bibinfo {year} {2015})}\BibitemShut {NoStop}%
\bibitem [{\citenamefont {Jagla}(2015)}]{jagla_avalanche-size_2015}%
  \BibitemOpen
  \bibfield  {author} {\bibinfo {author} {\bibfnamefont {E.~A.}\ \bibnamefont
  {Jagla}},\ }\href@noop {} {\bibfield  {journal} {\bibinfo  {journal}
  {Physical Review E}\ }\textbf {\bibinfo {volume} {92}},\ \bibinfo {pages}
  {042135} (\bibinfo {year} {2015})}\BibitemShut {NoStop}%
\bibitem [{\citenamefont {Liu}\ \emph {et~al.}(2016)\citenamefont {Liu},
  \citenamefont {Ferrero}, \citenamefont {Puosi}, \citenamefont {Barrat},\ and\
  \citenamefont {Martens}}]{liu_driving_2016}%
  \BibitemOpen
  \bibfield  {author} {\bibinfo {author} {\bibfnamefont {C.}~\bibnamefont
  {Liu}}, \bibinfo {author} {\bibfnamefont {E.~E.}\ \bibnamefont {Ferrero}},
  \bibinfo {author} {\bibfnamefont {F.}~\bibnamefont {Puosi}}, \bibinfo
  {author} {\bibfnamefont {J.-L.}\ \bibnamefont {Barrat}},\ and\ \bibinfo
  {author} {\bibfnamefont {K.}~\bibnamefont {Martens}},\ }\href@noop {}
  {\bibfield  {journal} {\bibinfo  {journal} {Physical review letters}\
  }\textbf {\bibinfo {volume} {116}},\ \bibinfo {pages} {065501} (\bibinfo
  {year} {2016})}\BibitemShut {NoStop}%
\bibitem [{\citenamefont {Lin}\ and\ \citenamefont
  {Wyart}(2016)}]{lin_mean-field_2016}%
  \BibitemOpen
  \bibfield  {author} {\bibinfo {author} {\bibfnamefont {J.}~\bibnamefont
  {Lin}}\ and\ \bibinfo {author} {\bibfnamefont {M.}~\bibnamefont {Wyart}},\
  }\href@noop {} {\bibfield  {journal} {\bibinfo  {journal} {Physical review
  X}\ }\textbf {\bibinfo {volume} {6}},\ \bibinfo {pages} {011005} (\bibinfo
  {year} {2016})}\BibitemShut {NoStop}%
\bibitem [{\citenamefont {Budrikis}\ \emph {et~al.}(2017)\citenamefont
  {Budrikis}, \citenamefont {Castellanos}, \citenamefont {Sandfeld},
  \citenamefont {Zaiser},\ and\ \citenamefont
  {Zapperi}}]{budrikis_universal_2017}%
  \BibitemOpen
  \bibfield  {author} {\bibinfo {author} {\bibfnamefont {Z.}~\bibnamefont
  {Budrikis}}, \bibinfo {author} {\bibfnamefont {D.~F.}\ \bibnamefont
  {Castellanos}}, \bibinfo {author} {\bibfnamefont {S.}~\bibnamefont
  {Sandfeld}}, \bibinfo {author} {\bibfnamefont {M.}~\bibnamefont {Zaiser}},\
  and\ \bibinfo {author} {\bibfnamefont {S.}~\bibnamefont {Zapperi}},\
  }\href@noop {} {\bibfield  {journal} {\bibinfo  {journal} {Nature
  communications}\ }\textbf {\bibinfo {volume} {8}},\ \bibinfo {pages} {15928}
  (\bibinfo {year} {2017})}\BibitemShut {NoStop}%
\bibitem [{\citenamefont {Fernández~Aguirre}\ and\ \citenamefont
  {Jagla}(2018)}]{fernandez_aguirre_critical_2018}%
  \BibitemOpen
  \bibfield  {author} {\bibinfo {author} {\bibfnamefont {I.}~\bibnamefont
  {Fernández~Aguirre}}\ and\ \bibinfo {author} {\bibfnamefont {E.~A.}\
  \bibnamefont {Jagla}},\ }\href {https://doi.org/10.1103/PhysRevE.98.013002}
  {\bibfield  {journal} {\bibinfo  {journal} {Physical Review E}\ }\textbf
  {\bibinfo {volume} {98}},\ \bibinfo {pages} {013002} (\bibinfo {year}
  {2018})},\ \bibinfo {note} {publisher: American Physical Society}\BibitemShut
  {NoStop}%
\bibitem [{\citenamefont {E. Ferrero}\ and\ \citenamefont
  {A. Jagla}(2019)}]{ferrero_criticality_2019}%
  \BibitemOpen
  \bibfield  {author} {\bibinfo {author} {\bibfnamefont {E.}~\bibnamefont
  {E. Ferrero}}\ and\ \bibinfo {author} {\bibfnamefont {E.}~\bibnamefont
  {A. Jagla}},\ }\href {https://doi.org/10.1039/C9SM01073D} {\bibfield
  {journal} {\bibinfo  {journal} {Soft Matter}\ }\textbf {\bibinfo {volume}
  {15}},\ \bibinfo {pages} {9041} (\bibinfo {year} {2019})},\ \bibinfo {note}
  {publisher: Royal Society of Chemistry}\BibitemShut {NoStop}%
\bibitem [{\citenamefont {Ferrero}\ and\ \citenamefont
  {Jagla}(2021)}]{ferrero_properties_2021}%
  \BibitemOpen
  \bibfield  {author} {\bibinfo {author} {\bibfnamefont {E.~E.}\ \bibnamefont
  {Ferrero}}\ and\ \bibinfo {author} {\bibfnamefont {E.~A.}\ \bibnamefont
  {Jagla}},\ }\href {https://doi.org/10.1088/1361-648X/abd73a} {\bibfield
  {journal} {\bibinfo  {journal} {Journal of Physics: Condensed Matter}\
  }\textbf {\bibinfo {volume} {33}},\ \bibinfo {pages} {124001} (\bibinfo
  {year} {2021})},\ \bibinfo {note} {publisher: IOP Publishing}\BibitemShut
  {NoStop}%
\bibitem [{\citenamefont {Lin}\ \emph {et~al.}(2014{\natexlab{b}})\citenamefont
  {Lin}, \citenamefont {Saade}, \citenamefont {Lerner}, \citenamefont {Rosso},\
  and\ \citenamefont {Wyart}}]{lin_density_2014}%
  \BibitemOpen
  \bibfield  {author} {\bibinfo {author} {\bibfnamefont {J.}~\bibnamefont
  {Lin}}, \bibinfo {author} {\bibfnamefont {A.}~\bibnamefont {Saade}}, \bibinfo
  {author} {\bibfnamefont {E.}~\bibnamefont {Lerner}}, \bibinfo {author}
  {\bibfnamefont {A.}~\bibnamefont {Rosso}},\ and\ \bibinfo {author}
  {\bibfnamefont {M.}~\bibnamefont {Wyart}},\ }\href@noop {} {\bibfield
  {journal} {\bibinfo  {journal} {EPL (Europhysics Letters)}\ }\textbf
  {\bibinfo {volume} {105}},\ \bibinfo {pages} {26003} (\bibinfo {year}
  {2014}{\natexlab{b}})}\BibitemShut {NoStop}%
\bibitem [{\citenamefont {Le~Goff}\ \emph {et~al.}(2019)\citenamefont
  {Le~Goff}, \citenamefont {Bertin},\ and\ \citenamefont
  {Martens}}]{le_goff_criticality_2019}%
  \BibitemOpen
  \bibfield  {author} {\bibinfo {author} {\bibfnamefont {M.}~\bibnamefont
  {Le~Goff}}, \bibinfo {author} {\bibfnamefont {E.}~\bibnamefont {Bertin}},\
  and\ \bibinfo {author} {\bibfnamefont {K.}~\bibnamefont {Martens}},\ }\href
  {https://doi.org/10.1103/PhysRevLett.123.108003} {\bibfield  {journal}
  {\bibinfo  {journal} {Physical Review Letters}\ }\textbf {\bibinfo {volume}
  {123}},\ \bibinfo {pages} {108003} (\bibinfo {year} {2019})},\ \bibinfo
  {note} {publisher: American Physical Society}\BibitemShut {NoStop}%
\bibitem [{\citenamefont {Karimi}\ \emph {et~al.}(2017)\citenamefont {Karimi},
  \citenamefont {Ferrero},\ and\ \citenamefont {Barrat}}]{karimi_inertia_2017}%
  \BibitemOpen
  \bibfield  {author} {\bibinfo {author} {\bibfnamefont {K.}~\bibnamefont
  {Karimi}}, \bibinfo {author} {\bibfnamefont {E.~E.}\ \bibnamefont
  {Ferrero}},\ and\ \bibinfo {author} {\bibfnamefont {J.-L.}\ \bibnamefont
  {Barrat}},\ }\href {https://doi.org/10.1103/PhysRevE.95.013003} {\bibfield
  {journal} {\bibinfo  {journal} {Physical Review E}\ }\textbf {\bibinfo
  {volume} {95}},\ \bibinfo {pages} {013003} (\bibinfo {year}
  {2017})}\BibitemShut {NoStop}%
\bibitem [{\citenamefont {Salerno}\ \emph {et~al.}(2012)\citenamefont
  {Salerno}, \citenamefont {Maloney},\ and\ \citenamefont
  {Robbins}}]{salerno_avalanches_2012}%
  \BibitemOpen
  \bibfield  {author} {\bibinfo {author} {\bibfnamefont {K.~M.}\ \bibnamefont
  {Salerno}}, \bibinfo {author} {\bibfnamefont {C.~E.}\ \bibnamefont
  {Maloney}},\ and\ \bibinfo {author} {\bibfnamefont {M.~O.}\ \bibnamefont
  {Robbins}},\ }\href {https://doi.org/10.1103/PhysRevLett.109.105703}
  {\bibfield  {journal} {\bibinfo  {journal} {Physical review letters}\
  }\textbf {\bibinfo {volume} {109}},\ \bibinfo {pages} {105703} (\bibinfo
  {year} {2012})}\BibitemShut {NoStop}%
\bibitem [{\citenamefont {Nicolas}\ \emph {et~al.}(2016)\citenamefont
  {Nicolas}, \citenamefont {Barrat},\ and\ \citenamefont
  {Rottler}}]{nicolas_effects_2016}%
  \BibitemOpen
  \bibfield  {author} {\bibinfo {author} {\bibfnamefont {A.}~\bibnamefont
  {Nicolas}}, \bibinfo {author} {\bibfnamefont {J.-L.}\ \bibnamefont
  {Barrat}},\ and\ \bibinfo {author} {\bibfnamefont {J.}~\bibnamefont
  {Rottler}},\ }\href {https://doi.org/10.1103/PhysRevLett.116.058303}
  {\bibfield  {journal} {\bibinfo  {journal} {Physical Review Letters}\
  }\textbf {\bibinfo {volume} {116}},\ \bibinfo {pages} {058303} (\bibinfo
  {year} {2016})}\BibitemShut {NoStop}%
\bibitem [{\citenamefont {Salerno}\ and\ \citenamefont
  {Robbins}(2013)}]{salerno_effect_2013}%
  \BibitemOpen
  \bibfield  {author} {\bibinfo {author} {\bibfnamefont {K.~M.}\ \bibnamefont
  {Salerno}}\ and\ \bibinfo {author} {\bibfnamefont {M.~O.}\ \bibnamefont
  {Robbins}},\ }\href {https://doi.org/10.1103/PhysRevE.88.062206} {\bibfield
  {journal} {\bibinfo  {journal} {Physical Review E}\ }\textbf {\bibinfo
  {volume} {88}},\ \bibinfo {pages} {062206} (\bibinfo {year} {2013})},\
  \bibinfo {note} {publisher: American Physical Society}\BibitemShut {NoStop}%
\bibitem [{\citenamefont {Nicolas}\ \emph {et~al.}(2018)\citenamefont
  {Nicolas}, \citenamefont {Ferrero}, \citenamefont {Martens},\ and\
  \citenamefont {Barrat}}]{nicolas_deformation_2018}%
  \BibitemOpen
  \bibfield  {author} {\bibinfo {author} {\bibfnamefont {A.}~\bibnamefont
  {Nicolas}}, \bibinfo {author} {\bibfnamefont {E.~E.}\ \bibnamefont
  {Ferrero}}, \bibinfo {author} {\bibfnamefont {K.}~\bibnamefont {Martens}},\
  and\ \bibinfo {author} {\bibfnamefont {J.-L.}\ \bibnamefont {Barrat}},\
  }\href@noop {} {\bibfield  {journal} {\bibinfo  {journal} {Reviews of Modern
  Physics}\ }\textbf {\bibinfo {volume} {90}},\ \bibinfo {pages} {045006}
  (\bibinfo {year} {2018})}\BibitemShut {NoStop}%
\bibitem [{\citenamefont {Homer}\ and\ \citenamefont
  {Schuh}(2009)}]{homer_mesoscale_2009}%
  \BibitemOpen
  \bibfield  {author} {\bibinfo {author} {\bibfnamefont {E.~R.}\ \bibnamefont
  {Homer}}\ and\ \bibinfo {author} {\bibfnamefont {C.~A.}\ \bibnamefont
  {Schuh}},\ }\href
  {https://doi.org/https://doi.org/10.1016/j.actamat.2009.02.035} {\bibfield
  {journal} {\bibinfo  {journal} {Acta Materialia}\ }\textbf {\bibinfo {volume}
  {57}},\ \bibinfo {pages} {2823} (\bibinfo {year} {2009})}\BibitemShut
  {NoStop}%
\bibitem [{\citenamefont {Shi}\ and\ \citenamefont
  {Falk}(2005)}]{shi_strain_2005}%
  \BibitemOpen
  \bibfield  {author} {\bibinfo {author} {\bibfnamefont {Y.}~\bibnamefont
  {Shi}}\ and\ \bibinfo {author} {\bibfnamefont {M.~L.}\ \bibnamefont {Falk}},\
  }\href {https://doi.org/10.1103/PhysRevLett.95.095502} {\bibfield  {journal}
  {\bibinfo  {journal} {Physical Review Letters}\ }\textbf {\bibinfo {volume}
  {95}},\ \bibinfo {pages} {095502} (\bibinfo {year} {2005})},\ \bibinfo {note}
  {publisher: American Physical Society}\BibitemShut {NoStop}%
\bibitem [{\citenamefont {Shi}\ and\ \citenamefont
  {Falk}(2006)}]{shi_atomic-scale_2006}%
  \BibitemOpen
  \bibfield  {author} {\bibinfo {author} {\bibfnamefont {Y.}~\bibnamefont
  {Shi}}\ and\ \bibinfo {author} {\bibfnamefont {M.~L.}\ \bibnamefont {Falk}},\
  }\href {https://doi.org/10.1103/PhysRevB.73.214201} {\bibfield  {journal}
  {\bibinfo  {journal} {Physical Review B}\ }\textbf {\bibinfo {volume} {73}},\
  \bibinfo {pages} {214201} (\bibinfo {year} {2006})},\ \bibinfo {note}
  {publisher: American Physical Society}\BibitemShut {NoStop}%
\bibitem [{\citenamefont {Karmakar}\ \emph {et~al.}(2010)\citenamefont
  {Karmakar}, \citenamefont {Lerner}, \citenamefont {Procaccia},\ and\
  \citenamefont {Zylberg}}]{karmakar_statistical_2010}%
  \BibitemOpen
  \bibfield  {author} {\bibinfo {author} {\bibfnamefont {S.}~\bibnamefont
  {Karmakar}}, \bibinfo {author} {\bibfnamefont {E.}~\bibnamefont {Lerner}},
  \bibinfo {author} {\bibfnamefont {I.}~\bibnamefont {Procaccia}},\ and\
  \bibinfo {author} {\bibfnamefont {J.}~\bibnamefont {Zylberg}},\ }\href
  {https://doi.org/10.1103/PhysRevE.82.031301} {\bibfield  {journal} {\bibinfo
  {journal} {Physical Review E}\ }\textbf {\bibinfo {volume} {82}},\ \bibinfo
  {pages} {031301} (\bibinfo {year} {2010})}\BibitemShut {NoStop}%
\bibitem [{\citenamefont {Bécu}\ \emph {et~al.}(2006)\citenamefont {Bécu},
  \citenamefont {Manneville},\ and\ \citenamefont
  {Colin}}]{becu_yielding_2006}%
  \BibitemOpen
  \bibfield  {author} {\bibinfo {author} {\bibfnamefont {L.}~\bibnamefont
  {Bécu}}, \bibinfo {author} {\bibfnamefont {S.}~\bibnamefont {Manneville}},\
  and\ \bibinfo {author} {\bibfnamefont {A.}~\bibnamefont {Colin}},\ }\href
  {https://doi.org/10.1103/PhysRevLett.96.138302} {\bibfield  {journal}
  {\bibinfo  {journal} {Physical Review Letters}\ }\textbf {\bibinfo {volume}
  {96}},\ \bibinfo {pages} {138302} (\bibinfo {year} {2006})},\ \bibinfo {note}
  {publisher: American Physical Society}\BibitemShut {NoStop}%
\bibitem [{\citenamefont {Caggioni}\ \emph {et~al.}(2020)\citenamefont
  {Caggioni}, \citenamefont {Trappe},\ and\ \citenamefont
  {Spicer}}]{caggioni_variations_2020}%
  \BibitemOpen
  \bibfield  {author} {\bibinfo {author} {\bibfnamefont {M.}~\bibnamefont
  {Caggioni}}, \bibinfo {author} {\bibfnamefont {V.}~\bibnamefont {Trappe}},\
  and\ \bibinfo {author} {\bibfnamefont {P.~T.}\ \bibnamefont {Spicer}},\
  }\href {https://doi.org/10.1122/1.5120633} {\bibfield  {journal} {\bibinfo
  {journal} {Journal of Rheology}\ }\textbf {\bibinfo {volume} {64}},\ \bibinfo
  {pages} {413} (\bibinfo {year} {2020})},\ \bibinfo {note} {publisher: The
  Society of Rheology}\BibitemShut {NoStop}%
\bibitem [{\citenamefont {Chattoraj}\ \emph {et~al.}(2010)\citenamefont
  {Chattoraj}, \citenamefont {Caroli},\ and\ \citenamefont
  {Lemaître}}]{chattoraj_universal_2010}%
  \BibitemOpen
  \bibfield  {author} {\bibinfo {author} {\bibfnamefont {J.}~\bibnamefont
  {Chattoraj}}, \bibinfo {author} {\bibfnamefont {C.}~\bibnamefont {Caroli}},\
  and\ \bibinfo {author} {\bibfnamefont {A.}~\bibnamefont {Lemaître}},\ }\href
  {https://doi.org/10.1103/PhysRevLett.105.266001} {\bibfield  {journal}
  {\bibinfo  {journal} {Physical Review Letters}\ }\textbf {\bibinfo {volume}
  {105}},\ \bibinfo {pages} {266001} (\bibinfo {year} {2010})},\ \bibinfo
  {note} {publisher: American Physical Society}\BibitemShut {NoStop}%
\bibitem [{\citenamefont {Popovi\ifmmode~\acute{c}\else \'{c}\fi{}}\ \emph
  {et~al.}(2021)\citenamefont {Popovi\ifmmode~\acute{c}\else \'{c}\fi{}},
  \citenamefont {de~Geus}, \citenamefont {Ji},\ and\ \citenamefont
  {Wyart}}]{popovic2021thermalflow}%
  \BibitemOpen
  \bibfield  {author} {\bibinfo {author} {\bibfnamefont {M.}~\bibnamefont
  {Popovi\ifmmode~\acute{c}\else \'{c}\fi{}}}, \bibinfo {author} {\bibfnamefont
  {T.~W.~J.}\ \bibnamefont {de~Geus}}, \bibinfo {author} {\bibfnamefont
  {W.}~\bibnamefont {Ji}},\ and\ \bibinfo {author} {\bibfnamefont
  {M.}~\bibnamefont {Wyart}},\ }\href
  {https://doi.org/10.1103/PhysRevE.104.025010} {\bibfield  {journal} {\bibinfo
   {journal} {Phys. Rev. E}\ }\textbf {\bibinfo {volume} {104}},\ \bibinfo
  {pages} {025010} (\bibinfo {year} {2021})}\BibitemShut {NoStop}%
\bibitem [{\citenamefont {Ferrero}\ \emph {et~al.}(2021)\citenamefont
  {Ferrero}, \citenamefont {Kolton},\ and\ \citenamefont
  {Jagla}}]{ferrero2021yielding}%
  \BibitemOpen
  \bibfield  {author} {\bibinfo {author} {\bibfnamefont {E.~E.}\ \bibnamefont
  {Ferrero}}, \bibinfo {author} {\bibfnamefont {A.~B.}\ \bibnamefont
  {Kolton}},\ and\ \bibinfo {author} {\bibfnamefont {E.~A.}\ \bibnamefont
  {Jagla}},\ }\href {https://doi.org/10.1103/PhysRevMaterials.5.115602}
  {\bibfield  {journal} {\bibinfo  {journal} {Phys. Rev. Materials}\ }\textbf
  {\bibinfo {volume} {5}},\ \bibinfo {pages} {115602} (\bibinfo {year}
  {2021})}\BibitemShut {NoStop}%
\bibitem [{\citenamefont {Popović}\ \emph {et~al.}(2021)\citenamefont
  {Popović}, \citenamefont {de~Geus}, \citenamefont {Ji}, \citenamefont
  {Rosso},\ and\ \citenamefont {Wyart}}]{popovic2021thermalcreep}%
  \BibitemOpen
  \bibfield  {author} {\bibinfo {author} {\bibfnamefont {M.}~\bibnamefont
  {Popović}}, \bibinfo {author} {\bibfnamefont {T.~W.~J.}\ \bibnamefont
  {de~Geus}}, \bibinfo {author} {\bibfnamefont {W.}~\bibnamefont {Ji}},
  \bibinfo {author} {\bibfnamefont {A.}~\bibnamefont {Rosso}},\ and\ \bibinfo
  {author} {\bibfnamefont {M.}~\bibnamefont {Wyart}},\ }\href
  {http://arxiv.org/abs/2111.04061} {\bibfield  {journal} {\bibinfo  {journal}
  {arXiv:2111.04061 [cond-mat]}\ } (\bibinfo {year} {2021})},\ \bibinfo {note}
  {arXiv: 2111.04061}\BibitemShut {NoStop}%
\bibitem [{\citenamefont {Bustingorry}\ \emph {et~al.}(2007)\citenamefont
  {Bustingorry}, \citenamefont {Kolton},\ and\ \citenamefont
  {Giamarchi}}]{bustingorry_thermal_2007}%
  \BibitemOpen
  \bibfield  {author} {\bibinfo {author} {\bibfnamefont {S.}~\bibnamefont
  {Bustingorry}}, \bibinfo {author} {\bibfnamefont {A.~B.}\ \bibnamefont
  {Kolton}},\ and\ \bibinfo {author} {\bibfnamefont {T.}~\bibnamefont
  {Giamarchi}},\ }\href {https://doi.org/10.1209/0295-5075/81/26005} {\bibfield
   {journal} {\bibinfo  {journal} {EPL (Europhysics Letters)}\ }\textbf
  {\bibinfo {volume} {81}},\ \bibinfo {pages} {26005} (\bibinfo {year}
  {2007})},\ \bibinfo {note} {publisher: IOP Publishing}\BibitemShut {NoStop}%
\bibitem [{\citenamefont {Maloney}\ and\ \citenamefont
  {Lacks}(2006)}]{maloney_energy_2006}%
  \BibitemOpen
  \bibfield  {author} {\bibinfo {author} {\bibfnamefont {C.~E.}\ \bibnamefont
  {Maloney}}\ and\ \bibinfo {author} {\bibfnamefont {D.~J.}\ \bibnamefont
  {Lacks}},\ }\href {https://doi.org/10.1103/PhysRevE.73.061106} {\bibfield
  {journal} {\bibinfo  {journal} {Physical Review E}\ }\textbf {\bibinfo
  {volume} {73}},\ \bibinfo {pages} {061106} (\bibinfo {year} {2006})},\
  \bibinfo {note} {publisher: American Physical Society}\BibitemShut {NoStop}%
\bibitem [{\citenamefont {Truhlar}\ \emph {et~al.}(1996)\citenamefont
  {Truhlar}, \citenamefont {Garrett},\ and\ \citenamefont
  {Klippenstein}}]{truhlar_current_1996}%
  \BibitemOpen
  \bibfield  {author} {\bibinfo {author} {\bibfnamefont {D.~G.}\ \bibnamefont
  {Truhlar}}, \bibinfo {author} {\bibfnamefont {B.~C.}\ \bibnamefont
  {Garrett}},\ and\ \bibinfo {author} {\bibfnamefont {S.~J.}\ \bibnamefont
  {Klippenstein}},\ }\href {https://doi.org/10.1021/jp953748q} {\bibfield
  {journal} {\bibinfo  {journal} {The Journal of Physical Chemistry}\ }\textbf
  {\bibinfo {volume} {100}},\ \bibinfo {pages} {12771} (\bibinfo {year}
  {1996})},\ \bibinfo {note} {publisher: American Chemical Society}\BibitemShut
  {NoStop}%
\bibitem [{\citenamefont {Nicolas}\ \emph {et~al.}(2014)\citenamefont
  {Nicolas}, \citenamefont {Rottler},\ and\ \citenamefont
  {Barrat}}]{nicolas_spatiotemporal_2014}%
  \BibitemOpen
  \bibfield  {author} {\bibinfo {author} {\bibfnamefont {A.}~\bibnamefont
  {Nicolas}}, \bibinfo {author} {\bibfnamefont {J.}~\bibnamefont {Rottler}},\
  and\ \bibinfo {author} {\bibfnamefont {J.-L.}\ \bibnamefont {Barrat}},\
  }\href {https://doi.org/10.1140/epje/i2014-14050-1} {\bibfield  {journal}
  {\bibinfo  {journal} {The European Physical Journal E}\ }\textbf {\bibinfo
  {volume} {37}},\ \bibinfo {pages} {50} (\bibinfo {year} {2014})}\BibitemShut
  {NoStop}%
\bibitem [{\citenamefont {Ruscher}\ and\ \citenamefont
  {Rottler}(2020)}]{ruscher_residual_2020}%
  \BibitemOpen
  \bibfield  {author} {\bibinfo {author} {\bibfnamefont {C.}~\bibnamefont
  {Ruscher}}\ and\ \bibinfo {author} {\bibfnamefont {J.}~\bibnamefont
  {Rottler}},\ }\href {https://doi.org/10.1039/D0SM01155J} {\bibfield
  {journal} {\bibinfo  {journal} {Soft Matter}\ }\textbf {\bibinfo {volume}
  {16}},\ \bibinfo {pages} {8940} (\bibinfo {year} {2020})},\ \bibinfo {note}
  {publisher: The Royal Society of Chemistry}\BibitemShut {NoStop}%
\bibitem [{\citenamefont {Liu}\ \emph {et~al.}(2021)\citenamefont {Liu},
  \citenamefont {Dutta}, \citenamefont {Chaudhuri},\ and\ \citenamefont
  {Martens}}]{liu_elastoplastic_2021}%
  \BibitemOpen
  \bibfield  {author} {\bibinfo {author} {\bibfnamefont {C.}~\bibnamefont
  {Liu}}, \bibinfo {author} {\bibfnamefont {S.}~\bibnamefont {Dutta}}, \bibinfo
  {author} {\bibfnamefont {P.}~\bibnamefont {Chaudhuri}},\ and\ \bibinfo
  {author} {\bibfnamefont {K.}~\bibnamefont {Martens}},\ }\href
  {https://doi.org/10.1103/PhysRevLett.126.138005} {\bibfield  {journal}
  {\bibinfo  {journal} {Physical Review Letters}\ }\textbf {\bibinfo {volume}
  {126}},\ \bibinfo {pages} {138005} (\bibinfo {year} {2021})},\ \bibinfo
  {note} {publisher: American Physical Society}\BibitemShut {NoStop}%
\bibitem [{\citenamefont {Tyukodi}\ \emph {et~al.}(2019)\citenamefont
  {Tyukodi}, \citenamefont {Vandembroucq},\ and\ \citenamefont
  {Maloney}}]{tyukodi_avalanches_2019}%
  \BibitemOpen
  \bibfield  {author} {\bibinfo {author} {\bibfnamefont {B.}~\bibnamefont
  {Tyukodi}}, \bibinfo {author} {\bibfnamefont {D.}~\bibnamefont
  {Vandembroucq}},\ and\ \bibinfo {author} {\bibfnamefont {C.~E.}\ \bibnamefont
  {Maloney}},\ }\href@noop {} {\bibfield  {journal} {\bibinfo  {journal}
  {Physical Review E}\ }\textbf {\bibinfo {volume} {100}},\ \bibinfo {pages}
  {043003} (\bibinfo {year} {2019})}\BibitemShut {NoStop}%
\bibitem [{\citenamefont {Parley}\ \emph {et~al.}(2020)\citenamefont {Parley},
  \citenamefont {Fielding},\ and\ \citenamefont {Sollich}}]{parley_aging_2020}%
  \BibitemOpen
  \bibfield  {author} {\bibinfo {author} {\bibfnamefont {J.~T.}\ \bibnamefont
  {Parley}}, \bibinfo {author} {\bibfnamefont {S.~M.}\ \bibnamefont
  {Fielding}},\ and\ \bibinfo {author} {\bibfnamefont {P.}~\bibnamefont
  {Sollich}},\ }\href {https://doi.org/10.1063/5.0033196} {\bibfield  {journal}
  {\bibinfo  {journal} {Physics of Fluids}\ }\textbf {\bibinfo {volume} {32}},\
  \bibinfo {pages} {127104} (\bibinfo {year} {2020})},\ \bibinfo {note}
  {publisher: American Institute of Physics}\BibitemShut {NoStop}%
\bibitem [{\citenamefont {Korchinski}\ \emph {et~al.}(2021)\citenamefont
  {Korchinski}, \citenamefont {Ruscher},\ and\ \citenamefont
  {Rottler}}]{korchinski_signatures_2021}%
  \BibitemOpen
  \bibfield  {author} {\bibinfo {author} {\bibfnamefont {D.}~\bibnamefont
  {Korchinski}}, \bibinfo {author} {\bibfnamefont {C.}~\bibnamefont
  {Ruscher}},\ and\ \bibinfo {author} {\bibfnamefont {J.}~\bibnamefont
  {Rottler}},\ }\href {https://doi.org/10.1103/PhysRevE.104.034603} {\bibfield
  {journal} {\bibinfo  {journal} {Physical Review E}\ }\textbf {\bibinfo
  {volume} {104}},\ \bibinfo {pages} {034603} (\bibinfo {year} {2021})},\
  \bibinfo {note} {publisher: American Physical Society}\BibitemShut {NoStop}%
\bibitem [{\citenamefont {Hentschel}\ \emph {et~al.}(2010)\citenamefont
  {Hentschel}, \citenamefont {Karmakar}, \citenamefont {Lerner},\ and\
  \citenamefont {Procaccia}}]{hentschel_size_2010}%
  \BibitemOpen
  \bibfield  {author} {\bibinfo {author} {\bibfnamefont {H.~G.~E.}\
  \bibnamefont {Hentschel}}, \bibinfo {author} {\bibfnamefont {S.}~\bibnamefont
  {Karmakar}}, \bibinfo {author} {\bibfnamefont {E.}~\bibnamefont {Lerner}},\
  and\ \bibinfo {author} {\bibfnamefont {I.}~\bibnamefont {Procaccia}},\ }\href
  {https://doi.org/10.1103/PhysRevLett.104.025501} {\bibfield  {journal}
  {\bibinfo  {journal} {Physical Review Letters}\ }\textbf {\bibinfo {volume}
  {104}},\ \bibinfo {pages} {025501} (\bibinfo {year} {2010})},\ \bibinfo
  {note} {publisher: American Physical Society}\BibitemShut {NoStop}%
\bibitem [{\citenamefont {Clemmer}\ \emph {et~al.}(2021)\citenamefont
  {Clemmer}, \citenamefont {Salerno},\ and\ \citenamefont
  {Robbins}}]{clemmer_criticality_2021}%
  \BibitemOpen
  \bibfield  {author} {\bibinfo {author} {\bibfnamefont {J.~T.}\ \bibnamefont
  {Clemmer}}, \bibinfo {author} {\bibfnamefont {K.~M.}\ \bibnamefont
  {Salerno}},\ and\ \bibinfo {author} {\bibfnamefont {M.~O.}\ \bibnamefont
  {Robbins}},\ }\href {https://doi.org/10.1103/PhysRevE.103.042605} {\bibfield
  {journal} {\bibinfo  {journal} {Physical Review E}\ }\textbf {\bibinfo
  {volume} {103}},\ \bibinfo {pages} {042605} (\bibinfo {year} {2021})},\
  \bibinfo {note} {arXiv: 2104.04620}\BibitemShut {NoStop}%
\bibitem [{\citenamefont {Lin}\ and\ \citenamefont
  {Wyart}(2018)}]{lin_microscopic_2018}%
  \BibitemOpen
  \bibfield  {author} {\bibinfo {author} {\bibfnamefont {J.}~\bibnamefont
  {Lin}}\ and\ \bibinfo {author} {\bibfnamefont {M.}~\bibnamefont {Wyart}},\
  }\href@noop {} {\bibfield  {journal} {\bibinfo  {journal} {Physical Review
  E}\ }\textbf {\bibinfo {volume} {97}},\ \bibinfo {pages} {012603} (\bibinfo
  {year} {2018})}\BibitemShut {NoStop}%
\bibitem [{\citenamefont {Gutowski}\ \emph {et~al.}(2012)\citenamefont
  {Gutowski}, \citenamefont {Lee}, \citenamefont {de~Bruyn},\ and\
  \citenamefont {Frisken}}]{gutowski_scaling_2012}%
  \BibitemOpen
  \bibfield  {author} {\bibinfo {author} {\bibfnamefont {I.~A.}\ \bibnamefont
  {Gutowski}}, \bibinfo {author} {\bibfnamefont {D.}~\bibnamefont {Lee}},
  \bibinfo {author} {\bibfnamefont {J.~R.}\ \bibnamefont {de~Bruyn}},\ and\
  \bibinfo {author} {\bibfnamefont {B.~J.}\ \bibnamefont {Frisken}},\ }\href
  {https://doi.org/10.1007/s00397-011-0614-6} {\bibfield  {journal} {\bibinfo
  {journal} {Rheologica Acta}\ }\textbf {\bibinfo {volume} {51}},\ \bibinfo
  {pages} {441} (\bibinfo {year} {2012})}\BibitemShut {NoStop}%
\bibitem [{\citenamefont {Lin}\ \emph {et~al.}(2015)\citenamefont {Lin},
  \citenamefont {Gueudré}, \citenamefont {Rosso},\ and\ \citenamefont
  {Wyart}}]{lin_criticality_2015}%
  \BibitemOpen
  \bibfield  {author} {\bibinfo {author} {\bibfnamefont {J.}~\bibnamefont
  {Lin}}, \bibinfo {author} {\bibfnamefont {T.}~\bibnamefont {Gueudré}},
  \bibinfo {author} {\bibfnamefont {A.}~\bibnamefont {Rosso}},\ and\ \bibinfo
  {author} {\bibfnamefont {M.}~\bibnamefont {Wyart}},\ }\href@noop {}
  {\bibfield  {journal} {\bibinfo  {journal} {Physical review letters}\
  }\textbf {\bibinfo {volume} {115}},\ \bibinfo {pages} {168001} (\bibinfo
  {year} {2015})}\BibitemShut {NoStop}%
\bibitem [{\citenamefont {Patinet}\ \emph {et~al.}(2016)\citenamefont
  {Patinet}, \citenamefont {Vandembroucq},\ and\ \citenamefont
  {Falk}}]{patinet_connecting_2016}%
  \BibitemOpen
  \bibfield  {author} {\bibinfo {author} {\bibfnamefont {S.}~\bibnamefont
  {Patinet}}, \bibinfo {author} {\bibfnamefont {D.}~\bibnamefont
  {Vandembroucq}},\ and\ \bibinfo {author} {\bibfnamefont {M.~L.}\ \bibnamefont
  {Falk}},\ }\href {https://doi.org/10.1103/PhysRevLett.117.045501} {\bibfield
  {journal} {\bibinfo  {journal} {Physical Review Letters}\ }\textbf {\bibinfo
  {volume} {117}},\ \bibinfo {pages} {045501} (\bibinfo {year} {2016})},\
  \bibinfo {note} {publisher: American Physical Society}\BibitemShut {NoStop}%
\bibitem [{\citenamefont {Barbot}\ \emph {et~al.}(2018)\citenamefont {Barbot},
  \citenamefont {Lerbinger}, \citenamefont {Hernandez-Garcia}, \citenamefont
  {García-García}, \citenamefont {Falk}, \citenamefont {Vandembroucq},\ and\
  \citenamefont {Patinet}}]{barbot_local_2018}%
  \BibitemOpen
  \bibfield  {author} {\bibinfo {author} {\bibfnamefont {A.}~\bibnamefont
  {Barbot}}, \bibinfo {author} {\bibfnamefont {M.}~\bibnamefont {Lerbinger}},
  \bibinfo {author} {\bibfnamefont {A.}~\bibnamefont {Hernandez-Garcia}},
  \bibinfo {author} {\bibfnamefont {R.}~\bibnamefont {García-García}},
  \bibinfo {author} {\bibfnamefont {M.~L.}\ \bibnamefont {Falk}}, \bibinfo
  {author} {\bibfnamefont {D.}~\bibnamefont {Vandembroucq}},\ and\ \bibinfo
  {author} {\bibfnamefont {S.}~\bibnamefont {Patinet}},\ }\href
  {https://doi.org/10.1103/PhysRevE.97.033001} {\bibfield  {journal} {\bibinfo
  {journal} {Physical Review E}\ }\textbf {\bibinfo {volume} {97}},\ \bibinfo
  {pages} {033001} (\bibinfo {year} {2018})},\ \bibinfo {note} {publisher:
  American Physical Society}\BibitemShut {NoStop}%
\bibitem [{\citenamefont {Virtanen}\ \emph {et~al.}(2020)\citenamefont
  {Virtanen}, \citenamefont {Gommers}, \citenamefont {Oliphant}, \citenamefont
  {Haberland}, \citenamefont {Reddy}, \citenamefont {Cournapeau}, \citenamefont
  {Burovski}, \citenamefont {Peterson}, \citenamefont {Weckesser},
  \citenamefont {Bright}, \citenamefont {{van der Walt}}, \citenamefont
  {Brett}, \citenamefont {Wilson}, \citenamefont {Millman}, \citenamefont
  {Mayorov}, \citenamefont {Nelson}, \citenamefont {Jones}, \citenamefont
  {Kern}, \citenamefont {Larson}, \citenamefont {Carey}, \citenamefont {Polat},
  \citenamefont {Feng}, \citenamefont {Moore}, \citenamefont {{VanderPlas}},
  \citenamefont {Laxalde}, \citenamefont {Perktold}, \citenamefont {Cimrman},
  \citenamefont {Henriksen}, \citenamefont {Quintero}, \citenamefont {Harris},
  \citenamefont {Archibald}, \citenamefont {Ribeiro}, \citenamefont
  {Pedregosa}, \citenamefont {{van Mulbregt}},\ and\ \citenamefont {{SciPy 1.0
  Contributors}}}]{2020SciPy-NMeth}%
  \BibitemOpen
  \bibfield  {author} {\bibinfo {author} {\bibfnamefont {P.}~\bibnamefont
  {Virtanen}}, \bibinfo {author} {\bibfnamefont {R.}~\bibnamefont {Gommers}},
  \bibinfo {author} {\bibfnamefont {T.~E.}\ \bibnamefont {Oliphant}}, \bibinfo
  {author} {\bibfnamefont {M.}~\bibnamefont {Haberland}}, \bibinfo {author}
  {\bibfnamefont {T.}~\bibnamefont {Reddy}}, \bibinfo {author} {\bibfnamefont
  {D.}~\bibnamefont {Cournapeau}}, \bibinfo {author} {\bibfnamefont
  {E.}~\bibnamefont {Burovski}}, \bibinfo {author} {\bibfnamefont
  {P.}~\bibnamefont {Peterson}}, \bibinfo {author} {\bibfnamefont
  {W.}~\bibnamefont {Weckesser}}, \bibinfo {author} {\bibfnamefont
  {J.}~\bibnamefont {Bright}}, \bibinfo {author} {\bibfnamefont {S.~J.}\
  \bibnamefont {{van der Walt}}}, \bibinfo {author} {\bibfnamefont
  {M.}~\bibnamefont {Brett}}, \bibinfo {author} {\bibfnamefont
  {J.}~\bibnamefont {Wilson}}, \bibinfo {author} {\bibfnamefont {K.~J.}\
  \bibnamefont {Millman}}, \bibinfo {author} {\bibfnamefont {N.}~\bibnamefont
  {Mayorov}}, \bibinfo {author} {\bibfnamefont {A.~R.~J.}\ \bibnamefont
  {Nelson}}, \bibinfo {author} {\bibfnamefont {E.}~\bibnamefont {Jones}},
  \bibinfo {author} {\bibfnamefont {R.}~\bibnamefont {Kern}}, \bibinfo {author}
  {\bibfnamefont {E.}~\bibnamefont {Larson}}, \bibinfo {author} {\bibfnamefont
  {C.~J.}\ \bibnamefont {Carey}}, \bibinfo {author} {\bibfnamefont
  {{\.I}.}~\bibnamefont {Polat}}, \bibinfo {author} {\bibfnamefont
  {Y.}~\bibnamefont {Feng}}, \bibinfo {author} {\bibfnamefont {E.~W.}\
  \bibnamefont {Moore}}, \bibinfo {author} {\bibfnamefont {J.}~\bibnamefont
  {{VanderPlas}}}, \bibinfo {author} {\bibfnamefont {D.}~\bibnamefont
  {Laxalde}}, \bibinfo {author} {\bibfnamefont {J.}~\bibnamefont {Perktold}},
  \bibinfo {author} {\bibfnamefont {R.}~\bibnamefont {Cimrman}}, \bibinfo
  {author} {\bibfnamefont {I.}~\bibnamefont {Henriksen}}, \bibinfo {author}
  {\bibfnamefont {E.~A.}\ \bibnamefont {Quintero}}, \bibinfo {author}
  {\bibfnamefont {C.~R.}\ \bibnamefont {Harris}}, \bibinfo {author}
  {\bibfnamefont {A.~M.}\ \bibnamefont {Archibald}}, \bibinfo {author}
  {\bibfnamefont {A.~H.}\ \bibnamefont {Ribeiro}}, \bibinfo {author}
  {\bibfnamefont {F.}~\bibnamefont {Pedregosa}}, \bibinfo {author}
  {\bibfnamefont {P.}~\bibnamefont {{van Mulbregt}}},\ and\ \bibinfo {author}
  {\bibnamefont {{SciPy 1.0 Contributors}}},\ }\href
  {https://doi.org/10.1038/s41592-019-0686-2} {\bibfield  {journal} {\bibinfo
  {journal} {Nature Methods}\ }\textbf {\bibinfo {volume} {17}},\ \bibinfo
  {pages} {261} (\bibinfo {year} {2020})}\BibitemShut {NoStop}%
\end{thebibliography}%

\end{document}